\documentclass[aps,notitlepage,nofootinbib,superscriptaddress]{revtex4-1}
\usepackage{epsfig}
\usepackage{amsmath}
\newcommand{\ben}{\begin{eqnarray}}
\newcommand{\een}{\end{eqnarray}}
\newcommand{\la}{\langle}
\newcommand{\ra}{\rangle}
\usepackage[active]{srcltx} % for kile....

\begin{document}
\title{Global fitting of single spin asymmetry: an attempt}

\author{Zhong-Bo Kang}
\email{zkang@bnl.gov}
\affiliation{RIKEN BNL Research Center,
                  Brookhaven National Laboratory,
                  Upton, NY 11973, USA}

\author{Alexei Prokudin}
\email{prokudin@jlab.org}
\affiliation{Theory Group,
                   Jefferson Lab, 
                   Newport News, VA 23606, USA}

\begin{abstract}
We present an attempt of global analysis of Semi-Inclusive Deep Inelastic Scattering (SIDIS) $\ell p^\uparrow \to \ell' \pi X$ data on single spin asymmetries and data on left-right asymmetry $A_N$ in $p^\uparrow p \to \pi X$ in order to simultaneously extract information on
Sivers function and twist-three quark-gluon Efremov-Teryaev-Qiu-Sterman (ETQS) function. We explore different possibilities such as node
of Sivers function in $x$ or $k_\perp$ in order to explain ``sign mismatch'' between these functions. We show that $\pi^\pm$ SIDIS data and 
$\pi^0$ STAR data can be well described in a combined TMD and twist-3 fit, however 
$\pi^\pm$ BRAHMS data are not described in a satisfactory way. This leaves open a question to the solution of the ``sign mismatch''. Possible
explanations are then discussed.  
\end{abstract}

\date{\today}
\maketitle

%%%%%%%%%%%%%%%%%%%%%%%%%%
\section{Introduction}
Single transverse spin asymmetries (SSAs) are a rich source of information on internal partonic structure of the nucleon \cite{Boer:2011fh}. The exploration of the underlying mechanisms has led us to realize that the SSAs are sensitive probes of the parton's transverse motion. There are two different yet related QCD factorization formalisms to incorparate such transverse components of the parton's momentum and to describe the observed asymmetries: the transverse momentum dependent (TMD) factorization and the collinear twist-three factorization approaches.

For processes such as single inclusive hadron production in proton-proton collisions, 
$p^\uparrow p\to hX$,
which exhibits only one characteristic hard scale, the transverse momentum $P_{h\perp}^2 \gg \Lambda_{\rm QCD}^2$ of the produced hadron, one could describe the SSAs in terms of twist-three quark-gluon correlation functions \cite{Efremov:1981sh,Efremov:1984ip,Qiu:1991pp,Qiu:1998ia,Koike:2009ge,Kang:2010zzb}. 
One of the well-known examples is the so-called Efremov-Teryaev-Qiu-Sterman (ETQS) function. 
Phenomenological extractions were performed in different papers \cite{Kouvaris:2006zy,Kanazawa:2011bg}. On the other hand, for processes such as Semi-Inclusive Deep Inelastic Scattering (SIDIS) which possesses two characteristic scales, photon's virtuality $Q$ and $P_{h\perp}$ of the produced hadron, one can use a TMD factorization formalism \cite{Ji:2004wu,Ji:2004xq} 
 in the region $ \Lambda_{\rm QCD}^2 < P_{h\perp}^2 \ll Q^2 $ and describe 
asymmetries with TMD functions. One of the most important TMDs is the Sivers function $f^{\perp q}_{1T}$ \cite{Sivers:1989cc,Sivers:1990fh} which describes 
$\sin(\phi_h -\phi_s)$ modulation in SIDIS on transversely polarized target \cite{Boer:1997nt}. Sivers functions have been extracted from SIDIS experimental data by various groups \cite{Efremov:2004tp,Vogelsang:2005cs,Anselmino:2005nn,Arnold:2008ap,Anselmino:2008sga}. 

These two formalisms are closely related to each other, and have been shown to be equivalent in the overlap region where both can apply \cite{Ji:2006ub,Koike:2007dg,Bacchetta:2008xw}. The relevant functions - the Sivers function and the ETQS twist-three function - are connected through the following relation \cite{Boer:2003cm,Ma:2003ut,Kang:2011hk}\footnote{Note that this relation holds for the 
so-called unsubtracted part
of TMDs, see Ref.~\cite{collins}.}:
\ben
T_{q,F}(x,x) = - \int d^2 k_\perp \frac{|k_\perp^2|}{M} f^{\perp q}_{1T}(x,k_\perp^2)|_{\rm SIDIS}\; ,
\label{eq:relation}
\een
where the subscript ``SIDIS'' emphasizes that the Sivers function is probed in the SIDIS process. Color gauge invariant nature
of TMDs manifests itself in the fact that TMDs are process dependent  and an important consequence of this process-dependence is a prediction \cite{Collins:2002kn}:
\ben
f^{\perp q}_{1T}|_{\rm SIDIS} = - f^{\perp q}_{1T}|_{\rm DY}\; .
\label{eq:sivers}
\een
i.e., the Sivers function measured in SIDIS and Drell-Yan (DY) processes are exactly opposite to each other. Experiments are actively planning to measure and verify such a prediction. Some preliminary phenomenological estimates on the SSAs of DY production \cite{Collins:2005rq,Kang:2009sm, Anselmino:2009st} and solid theoretical developments \cite{Kang:2011mr,Aybat:2011ge} have been achieved.

Recently it was found that left-hand-side (LHS) and right-hand-side (RHS) of Eq.~\eqref{eq:relation} have opposite signs if the corresponding functions are extracted from phenomenological studies of different experimental data \cite{Kang:2011hk}, particularly the RHS $f^{\perp q}_{1T}(x,k_\perp^2)$ from the SIDIS data, while LHS $T_{q,F}(x,x)$ from $pp$ data. We will refer this finding as ``sign puzzle'' or ``sign mismatch''. Whether it reflects the incompatibility of SIDIS and $pp$ data within the current theoretical formalism, or reflects the inconsistency of our formalism itself, is a very important question and needs to be further explored both theoretically and experimentally. On the experimental side, the measurement for the SSAs of single inclusive jet and direct photon production \cite{Kang:2011hk}, and the single lepton production from a $W$-boson decay \cite{Metz:2010xs,Kang:2009bp} in $pp$ collisions, the single inclusive jet and hadron production in $\ell p$ collisions (without identifying the final-state lepton) \cite{Anselmino:2009pn,Kang:2011jw,Contalbrigo:2010zz} could be very helpful. 
The study of hadron distribution inside a jet could also be useful \cite{Yuan:2007nd,D'Alesio:2010am,D'Alesio:2011mc,Fersch:2011zz}.

In this paper we will try to make some first attempt on the theoretical (phenomenological) side: we will attempt to make a global fitting of both SIDIS and $pp$ data with a more flexible functional form for the Sivers function (and the ETQS function), to see if we are able to describe all the data within our current theoretical formalism. Our naive starting point is based on the observation that the SIDIS and $pp$ data typically covers slightly different kinematic region, in either momentum fraction $x$ and/or the transverse components. Thus a sign-changing functional form in these kinematic space might be just needed to cure the ``sign mismatch''. One such possibility, e.g., a node in $x$ region has already been indicated in Ref.~\cite{Boer:2011fx}. We will consider SIDIS data from HERMES and COMPASS, and proton-proton data from STAR and BRAHMS. Let us emphasize that this is a first attempt to use simultaneously the TMD and collinear twist-three factorization formalisms in a global analysis of the spin asymmetry. 

The rest of our paper is organized as follows. In Sec.~II we recall the basic formalisms needed to describe SIDIS data for semi-inclusive hadron production at low $P_{h\perp}$, and proton-proton data for inclusive hadron production at high $P_{h\perp}$. In Sec.~III we first introduce our more flexible parametrized functional form for the Sivers function, describe our fitting procedure. Particularly we explore the possibility of a node in $x$ region, and investigate whether it can help resolve the ``sign mismatch'' problem. At the end of this section we briefly comment on the possibility of the node in $k_\perp$ space. We conclude our paper in Sec.~IV.

%%%%%%%%%%%%%%%%%%%%%%%%%%%%
\section{Basic TMD and collinear twist-3 factorization formalisms}
In this section we review the basic formulas for the spin asymmetries in both SIDIS and proton-proton processes. We start with the semi-inclusive hadron production at low $P_{h\perp}$ in SIDIS, $e (\ell) + A^\uparrow(P, s_\perp)\to e (\ell') + h(P_h) + X$, which can be described by the TMD factorization formalism. The differential cross section for the so-called Sivers effect reads \cite{Bacchetta:2006tn}:
\ben
\frac{d\sigma}{dx_B \, dy\, d\phi_s \,dz_h\, d\phi_h\, P_{h\perp} d P_{h\perp}}
&=& \sigma_0(x_B, y, Q^2)
\left[F_{UU ,T} +  |{\bf s}_\perp|\, 
  \sin(\phi_h-\phi_s)\,
F_{UT ,T}^{\sin\left(\phi_h -\phi_s\right)} \right],
  \label{eq:aut}
\een
where $\sigma_0 = \frac{\alpha^2}{x_B y\, Q^2}\left(1+(1-y)^2\right)$ with $\alpha$ the fine structure constant, $q=\ell-\ell'$ with $q^2=-Q^2$, and the usual SIDIS variables are defined as
\ben
x_B=\frac{Q^2}{2P\cdot q},\qquad
y=\frac{P\cdot q}{P\cdot \ell},\qquad 
z_h=\frac{P\cdot P_h}{P\cdot q}.
\een
The Sivers asymmetry can be defined as the $\sin(\phi_h -\phi_s)$ modula:
\ben
A_{UT}^{\sin(\phi_h -\phi_s)} \equiv 2 \la  \sin(\phi_h -\phi_s) \ra  = 
2\frac{\int d \phi_h d \phi_s \sin(\phi_h -\phi_s) \left(d\sigma(\phi_h,\phi_s) - d\sigma(\phi_h,\phi_s+\pi)\right)}{\int d \phi_h d \phi_s \left(d\sigma(\phi_h,\phi_s) + d\sigma(\phi_h,\phi_s+\pi)\right)}\;,
\label{eq:prokudin_ssa}
\een
where the subscript $U$ stands for unpolarized lepton beam, and $T$ for the transverse polarization of the target nucleon. In terms of structure functions one has
\ben
A_{UT}^{\sin(\phi_h -\phi_s)}(x_B, z_h, P_{h\perp})   =  \frac{ \sigma_0(x_B, y, Q^2)
 }{ \sigma_0(x_B, y, Q^2)}\frac{F_{UT,T}^{\sin(\phi_h -\phi_s)}}{ F_{UU ,T}}
\label{eq:prokudin_ssa_sivers}.
\een
The structure functions depend on $x_B$, $Q^2$, $z_h$ and $P_{h\perp}^2$, and can be written as \cite{Mulders:1995dh,Bacchetta:2006tn}
\ben
F_{UU,T}\;
&=& x_B\,
\sum_a e_a^2 \int d^2 {\bf k}_\perp\,  f_{a/A}(x_B,k_\perp^2)\,D_{h/a}\left(z_h,({\bf P}_{h \perp} - z_h {\bf k}_\perp)^2\right),
\\
F_{UT,T}^{\sin(\phi_h -\phi_s)}\;
 &=& -x_B\,
\sum_a e_a^2 \int d^2 {\bf k}_\perp\, \frac{{\bf \hat h}\cdot{\bf k}_\perp}{M} f_{1T}^{\perp a}(x_B, k_\perp^2)\,D_{h/a}\left(z_h, ({\bf P}_{h \perp} - z_h {\bf k}_\perp)^2\right)\; ,
\label{eq:prokudin_structure_functions}
\een
where ${\bf \hat h}\equiv {\bf P}_{h \perp}/|{\bf P}_{h \perp}|$, $f_{1T}^{\perp a}$ is the Sivers function, $f_{a/A}$ and $D_{h/a}$ are TMD parton distribution function (PDF) and fragmentation function (FF), respectively. All our definitions on the TMD functions and these expressions are consistent with the Trento convention \cite{Bacchetta:2004jz}, which have been used in the experiments \cite{:2009ti,:2008dn}.

On the other hand, for single inclusive hadron production at high $P_{h\perp}$ in $p^\uparrow p$ collisions, $A^\uparrow(P, s_\perp)+B(P')\to h(P_h)+X$, the spin-averaged cross section $d\sigma\equiv [d\sigma(s_\perp)+d\sigma(-s_\perp)]/2$ is usually written in the collinear factorization formalism as,
\ben
E_h\frac{d\sigma}{d^3 P_h}&=&
\frac{\alpha_s^2}{S}\sum_{a,b,c} \int \frac{dz}{z^2} D_{h/c}(z)
\int \frac{dx'}{x'}f_{b/B}(x')\int \frac{dx}{x}
f_{a/A}(x)
H^U_{ab\to c}(\hat s,\hat t,\hat u)\delta\left(\hat s+\hat t+\hat u\right),
\label{avg}
\een
where $\sum_{a,b,c}$ runs over all parton flavors, $S=(P+P')^2$, $f_{a/A}(x)$ and $f_{b/B}(x')$ are the collinear PDFs, and $D_{h/c}(z)$ is the collinear FF. $H^U_{ab\to c}$ are the well-known unpolarized hard-part functions for partonic scattering \cite{Owens:1986mp,Kang:2010vd}. 
$\hat s$, $\hat t$, and $\hat u$ are the usual partonic Mandelstam variables, for the final hadron of transverse momentum $P_{h\perp}$ and rapidity $y$ we obtain
\ben
\hat s=x x' S,
\qquad
\hat t= - x/zP_{h\perp} \sqrt{S} e^{-y},
\qquad
\hat u= - x' /z P_{h\perp} \sqrt{S} e^{y}.
\een
The commonly used Feynman-$x_F$ can be written as $x_F=\frac{2P_{h\perp}}{\sqrt{S}} \sinh(y)$. Note that the partonic $x,x'$ and $z$ are integrated over in Eq.~\eqref{avg}.

The spin-dependent cross section $d\Delta\sigma(s_\perp)\equiv [d\sigma(s_\perp)-d\sigma(-s_\perp)]/2$ is given by the collinear twist-three factorization formalism:
\ben
\left. E_h\frac{d\Delta\sigma(s_\perp)}{d^3 P_h}\right|_{\rm Sivers}&=&
\epsilon_{\alpha \beta} s_\perp^\alpha P_{h\perp}^\beta
\frac{\alpha_s^2}{S}\sum_{a,b,c} \int \frac{dz}{z^2} D_{h/c}(z)
\int \frac{dx'}{x'}f_{b/B}(x')\int \frac{dx}{x}
\left[\frac{1}{z \hat{u}}\right]
\nonumber \\
&&
\times
\left[T_{a,F}(x, x) - x\frac{d}{dx}T_{a,F}(x, x)\right]
H^{\rm Sivers}_{ab\to c}(\hat s,\hat t,\hat u)\delta\left(\hat s+\hat t+\hat u\right),
\label{aftera}
\een
where $\epsilon_{\alpha\beta}$ is a two-dimensional anti-symmetric tensor with $\epsilon_{12}=1$ (and $\epsilon_{21}=-1$), $T_{a,F}(x, x)$ is twist-tree ETQS functions, and $H^{\rm Sivers}_{ab\to c}(\hat s,\hat t,\hat u)$ are the relevant hard-part functions which have been given in Refs.~\cite{Kouvaris:2006zy,Gamberg:2010tj}. The subscript ``Sivers'' here is to remind that there are other types of contributions to the SSAs for the inclusive hadron production. What is written in Eq.~\eqref{aftera} is only the so-called soft gluon pole contribution \cite{Kouvaris:2006zy}, and there could be soft fermion pole contribution \cite{Koike:2009ge}, and also the contribution from the twist-three fragmentation function \cite{Kang:2010zzb}. Nevertheless, the extensive phenomenological study of the single inclusive hadron production has been performed for the soft gluon pole contribution \cite{Kouvaris:2006zy,Kanazawa:2011bg}, which indicates the soft fermion pole contribution is relatively small at least in the forward region where the asymmetry is the largest \cite{Kanazawa:2011bg}. Our study in the current paper will also concentrate on the soft gluon pole contribution, for which the relevant twist-three function - ETQS function $T_{a,F}(x, x)$ - has a close relation to the Sivers function as in Eq.~\eqref{eq:relation} and thus we are able to perform a global analysis for both SIDIS and proton-proton data. We will comment on the contribution of the twist-three fragmentation function in the end of the next section.

The SSA, $A_N$, is given by the ratio of spin-dependent and spin-averaged cross sections
\ben
A_N\equiv\left.E_h\frac{d\Delta\sigma(s_\perp)}{d^3 P_h}\right/E_h\frac{d\sigma}{d^3 P_h},
\label{andef}
\een
The absolute sign of $A_N$ depends on the choice of frame and the coordinate system.  In the center-of-mass frame of the incoming hadrons $A$ and $B$, a convenient coordinate system (consistent with the experimental convention) is: the polarized nucleon $A$ moves along $+z$, unpolarized $B$ along $-z$, spin $s_\perp$ along $y$, and transverse momentum $P_{h\perp}$ along $x$-direction, respectively. In this frame
\ben
\epsilon_{\alpha \beta}s_\perp^\alpha P_{h\perp}^\beta=-P_{h\perp},
\label{epsilon}
\een
which should be used in Eq.~\eqref{aftera}.

%%%%%%%%%%%%%%%%%%%%%%%%%%%%
\section{Global fit of the spin asymmetry: an attempt}
So far all the phenomenological studies on the spin asymmetries in the market have been separated into two isolated parts. On one side, people use TMD factorization formalism to describe the SIDIS data for hadron production at low $P_{h\perp}$, and they solely concentrate on SIDIS data, and do not include proton-proton data in the global fitting. The Sivers functions have been extracted as a result of such studies. On the other side, collinear twist-3 factorization formalism is used to describe the proton-proton data for single inclusive hadron production at high $P_{h\perp}$, and only proton-proton data are analyzed without inclusion of SIDIS data in the global fitting. The so-called ETQS functions have been extracted from such studies. However, as we emphasize in our introduction, Sivers and ETQS functions are closely related. Thus in this section we attempt to perform a global analysis of both SIDIS and proton-proton data on the spin asymmetries. We will use TMD formalism to describe the SIDIS data in terms of the Sivers function. From the parametrization of the Sivers function, we obtain the functional form for ETQS function through Eq.~\eqref{eq:relation}. Then we use the collinear twist-3 formalism to describe proton-proton data in terms of our obtained ETQS function. In this way, we hope a single parameterization for the Sivers function could help us achieve a global fitting of both SIDIS and proton-proton data. We first introduce our parametrization for both Sivers and ETQS functions, then we present and discuss the results from our global fitting. We explore the possibility of node in $x$ in details, and briefly comment on the possibility of node in $k_\perp$ at the end of this section.

\subsection{Parametrization for the Sivers and ETQS function}
Following Refs.~\cite{Anselmino:2005nn,Anselmino:2008sga}, we parametrize both the spin-averaged PDF $f_{a/A}(x, k_\perp^2)$ and FF $D_{h/a}(z, p_T^2)$ with a Gaussian form for the transverse components:
\ben
f_{a/A}(x, k_\perp^2)&=&f_{a/A}(x) \frac{1}{\pi \la  k_\perp^2\ra }e^{-\mathbf{k}_\perp^2/\la  k_\perp^2\ra },
\\
D_{h/a}(z, p_{T}^2) &=& D_{h/a}(z)\,\frac{1}{\pi \la p_T^2 \ra} e^{-\mathbf{p}_{T}^2/\la p_T^2 \ra},
\een
such that they reduce to the usual collinear PDF $f_{a/A}(x)$ and FF $D_{h/a}(z)$ once integrated over the transverse momentum. The Gaussian widths $\la  k_\perp^2\ra =0.25$ GeV$^2$ and $\la p_T^2 \ra=0.20$ GeV$^2$ \cite{Anselmino:2005nn}.
The Sivers function $f_{1T}^{\perp q}(x,k_\perp^2)$ in SIDIS process will be parametrized as
\ben
f_{1T}^{\perp q}(x, k_\perp^2) &=& - \, {\cal N}_q(x)  h(k_\perp) \,
f_{q/A} (x, k_\perp^2), 
\label{eq:sivfac}
\een
where the extra $k_\perp$-dependence $h(k_\perp)$ is given by
\ben
h(k_\perp) = \sqrt{2e}\,\frac{M}{M_{1}}\,e^{-\mathbf{k}_\perp^2/{M_{1}^2}},
\label{eq:siverskt}
\een
with $M$ the nucleon mass, and $M_1$ a fitting parameter. The $x$-dependent part ${\cal N}_q(x)$ will be parametrized as
\ben
{\cal N}_q(x) =  N_q x^{\alpha_q}(1-x)^{\beta_q}
\frac{(\alpha_q+\beta_q)^{(\alpha_q+\beta_q)}}
{\alpha_q^{\alpha_q} \beta_q^{\beta_q}} (1-\eta_q x),
\label{eq:siversx}
\een
Compared with the previous SIDIS fits in Refs.~\cite{Anselmino:2005nn,Anselmino:2008sga}, the new ingredient lies in the factor $(1-\eta_q x)$, which is inspired from DSSV global fitting for the helicity PDFs~\cite{deFlorian:2008mr,deFlorian:2009vb}. This is a simplest form that can allow a node in the $x$ space: if $\eta_q>1$, we will have a node for $x\in [0,1]$; on the contrary, if $\eta_q<1$, then no node in the region $x\in [0,1]$.  
In our fit, to satisfy the positivity bound for Sivers function, we have to require $|{\cal N}_q(x)|<{1}$. To achieve this, we make the following substitution:
\ben
N_q \rightarrow N_q/{\rm max}\{1, |1-\eta_q|\} \; ,
\label{eq:nq}
\een
in Eq.~\eqref{eq:siversx} and allow $N_q$ to vary only inside the range $[-1,1]$, this enforces the positivity of Sivers function in $x\in [0,1]$.

Now through the relation between $T_{q,F}(x, x)$ and the Sivers function $f_{1T}^{\perp q}(x, k_\perp^2)$ in Eq.~\eqref{eq:relation}, we could thus obtain a parameterized form for $T_{q,F}(x, x)$ as
\ben
T_{q,F}(x, x) = \frac{\sqrt{2e}\la k_\perp^2\ra M_1^3}{\left(\la k_\perp^2\ra+M_1^2\right)^2} 
{\cal N}_q(x) f_{q/A} (x).
\een
In other words, once a parametrization of the quark Sivers function is given, we automatically have a parametrized form for ETQS matrix element $T_{q,F}(x, x)$. With this in hand, one will be able to make a simultaneous fit of both SIDIS at low $P_{h\perp}$ data and $pp$ inclusive hadron production at high $P_{h\perp}$. As a first attempt, we will only consider $u$ and $d$ quark flavors, and include only pion data ($\pi^{\pm,0}$) in our fit. Thus we have $N_q$, $\alpha_q$, $\beta_q$, and $\eta_q$ for both $u$ and $d$ quarks, and $M_1$ as our parameters, in total 9 parameters, to be determined by fitting the experimental data.

%%%%%%%%%%%%%%%%%%%%%%%%%%%%%%%%%%%  
\subsection{Description of the data and discussion}
As we have emphasized in last subsection, the TMD factorization formula of Eq.~\eqref{eq:prokudin_ssa_sivers} will be used to describe HERMES experiment proton target data \cite{:2009ti} and COMPASS experiments data \cite{:2008dn} on deuteron target. The twist-3 factorization formula Eqs.~(\ref{avg},\ref{aftera},\ref{andef}) will be used for $A_N$ data from STAR \cite{:2008qb} and BRAHMS \cite{Lee:2007zzh} experiments.  We use GRV98LO for the unpolarized PDFs \cite{Gluck:1998xa}, and DSS parametrization for the unpolarized FFs \cite{deFlorian:2007aj}. In our theoretical formalism we choose the factorization scale to be equal to the renormalization scale: $\mu=Q$ for SIDIS and $\mu=P_{h\perp}$ for proton-proton data.

\begin{figure*}[hbt]
\centering
  \begin{tabular}{c@{\hspace*{5mm}}c}
    \includegraphics[scale=0.4]{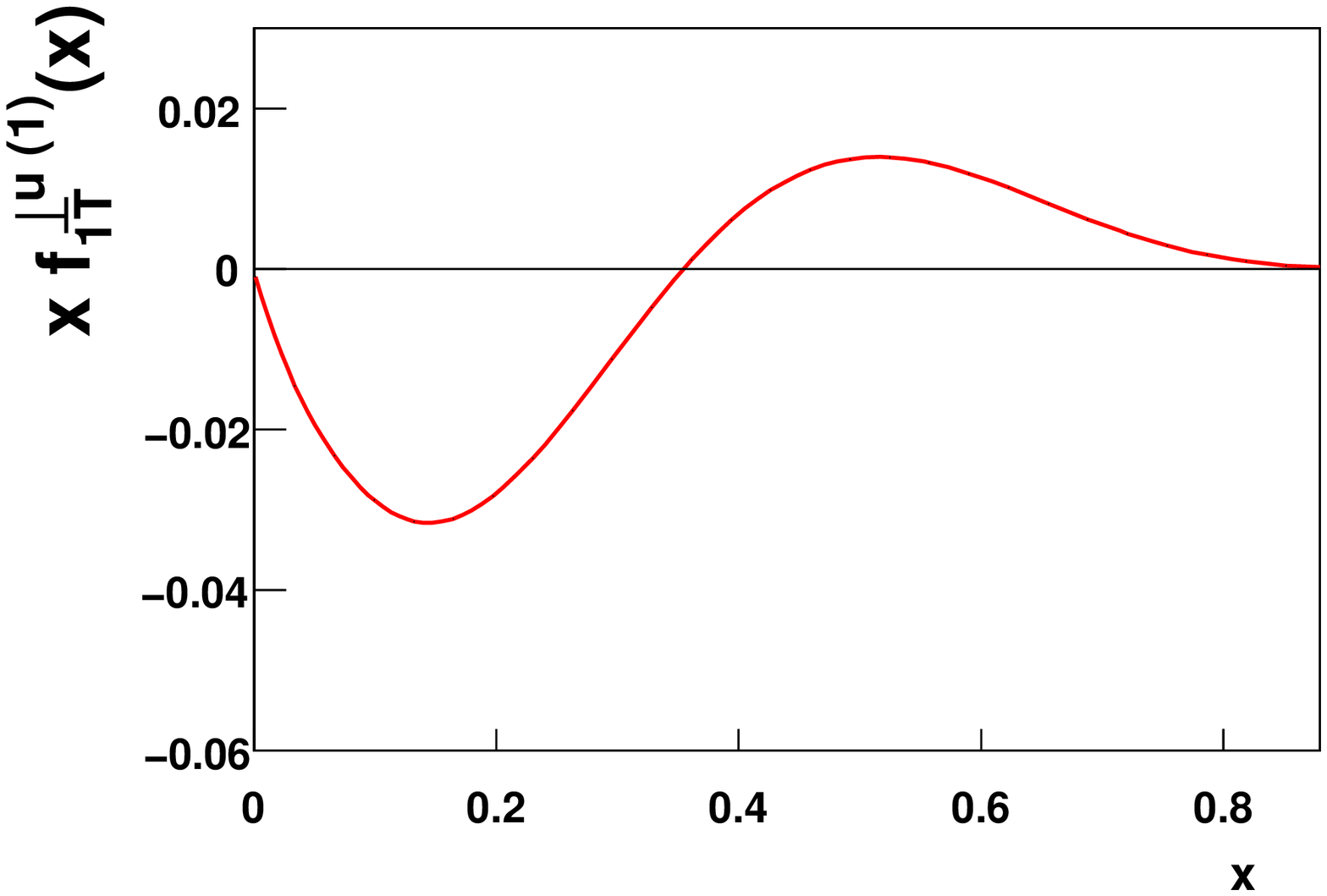}
    &
    \includegraphics[scale=0.4]{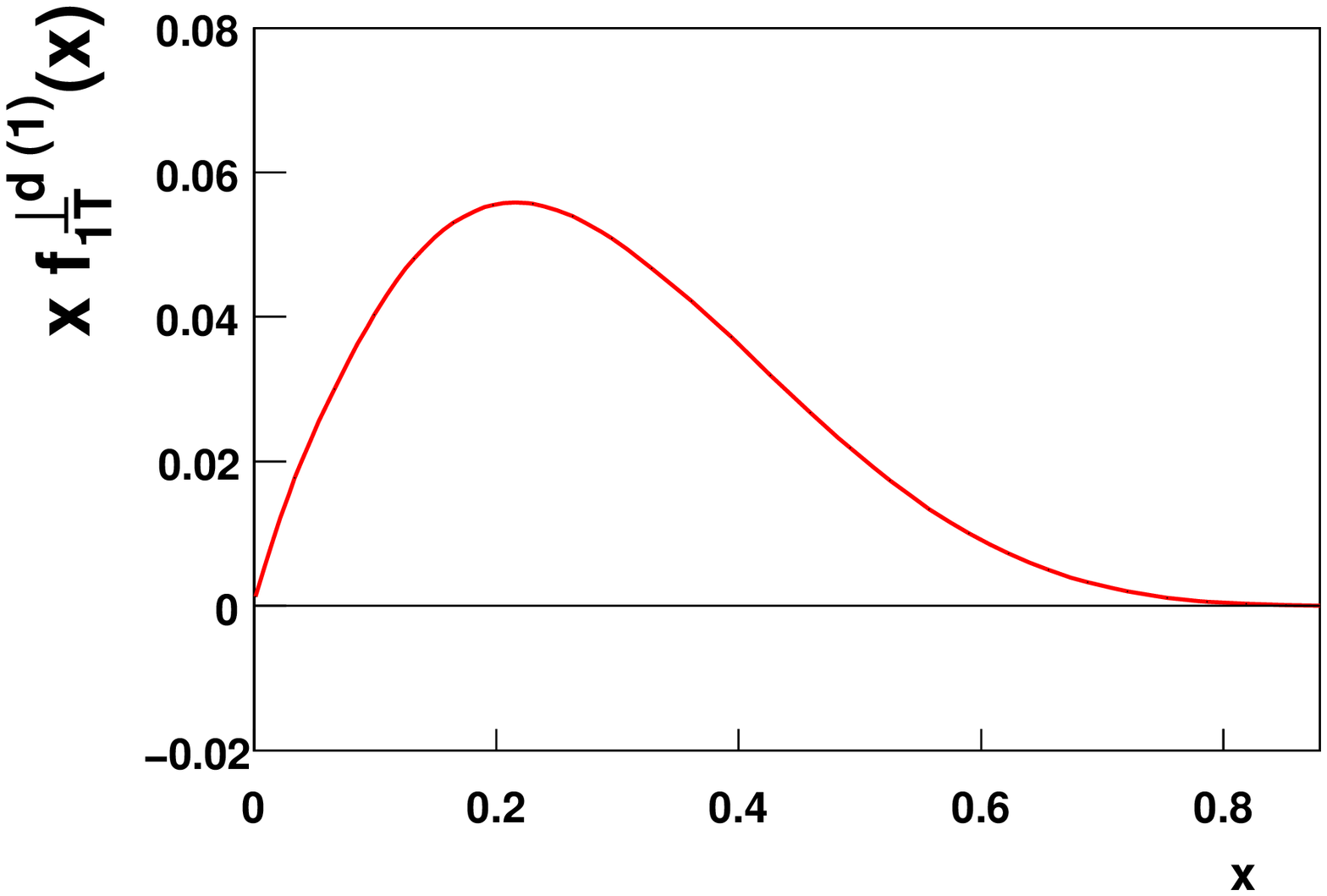}
  \\
  (a) & (b)
  \end{tabular}
\caption{ First moment of (a) $u$ quark Sivers function  and (b)  $d$ quark Sivers function as found using parameters of Table~\ref{table:I}, here $f_{1T}^{\perp q(1)}(x)\equiv -T_{q,F}(x,x)/2M$.}
\label{fig:sivers}
\end{figure*}

The results we obtain for the 9 free parameters by fitting simultaneously HERMES and COMPASS data sets on the Sivers asymmetry $A_{UT}^{\sin(\phi_h -\phi_s)}$, and the STAR and BRAHMS data sets on the SSAs $A_N$ for both charged and neutral pions, are presented in Table~\ref{table:I}. 
\begin{table}[t]
\caption{Best values of the free parameters for the $u$ and $d$
Sivers distribution functions, as obtained by 
simultaneously fitting HERMES \cite{:2009ti} and COMPASS \cite{:2008dn}  data on the 
$A_{UT}^{\sin(\phi_h-\phi_s)}$ asymmetry and the STAR \cite{:2008qb} and BRAHMS \cite{Lee:2007zzh} data on the $A_{N}$ 
asymmetry.
\label{table:I}}
\begin{center}
\begin{tabular}{lllllll}
\hline
\hline
  ~& ~& $\chi ^2/{\rm d.o.f.}\;\;$ = & $3.6$ & ~ & ~\\
\hline
$N_{u}$ &=&  $1$ ~~~& $N_{d}$ &=&  $\!-1$ \\
$\alpha_u$ &=&  $0.8$ & $\alpha_d$ &=&  $0.8$   \\
$\beta_u$  &=&  $1.5$  & $\beta_d$  &=&  $1$   \\
$\eta_u$ &=&  $2.8$   & $\eta_d$  &=&  $0$\\
$M_1^2$ &=&  0.7 GeV$^2$ & & &  \\
\hline
\hline
\end{tabular}
\end{center}
\end{table}
The extracted first moments of the Sivers functions for both $u$ and $d$ quark flavors are plotted in Fig.~\ref{fig:sivers}. One can see that $\eta_d=0$ thus no node for $d$ quark Sivers function. While
$\eta_u = 2.8>1$, there is a node for $u$-quark Sivers function that is located at $x_{\rm node} = 0.36$. This value is at the border of the region probed by SIDIS experiments and thus in principle the node cannot be excluded. $N_u=1$, $N_{d}=-1$ and thus the positivity bound is saturated by these functions at least at some point of $x$ value\footnote{Since $N_u=1$ and $|1-\eta_u|>1$, we have $|{\cal N}_u(x)|<1$ in Eqs.~(\ref{eq:siversx},\ref{eq:nq}) for the whole $x$ region except for $x=1$ where the bound is saturated, but both Sivers function and the unpolarized PDF vanish, and thus the positivity bound is satisfied for $u$-quark Sivers function. On the other hand, $N_{d}=-1$ and at the same time $\eta_d=0$, checking ${\cal N}_d(x)$ in Eqs.~(\ref{eq:siversx},\ref{eq:nq}) we find that the positivity bound could be saturated at some point of $x\neq 0$ for $d$-quark Sivers function.}.

\begin{figure*}[hbt]
\centering
  \begin{tabular}{c@{\hspace*{5mm}}c}
    \includegraphics[scale=0.4]{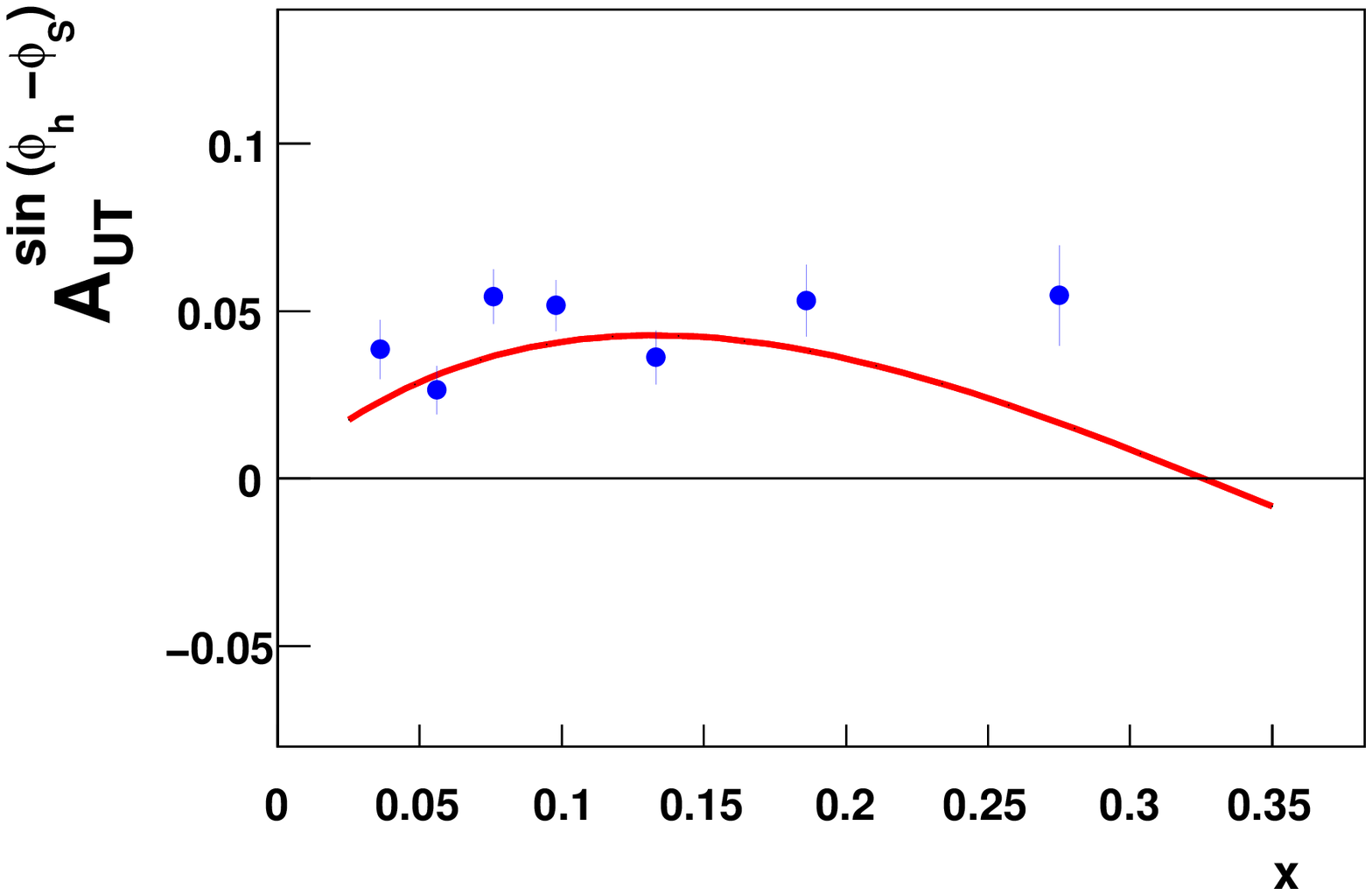}
    &
    \includegraphics[scale=0.4]{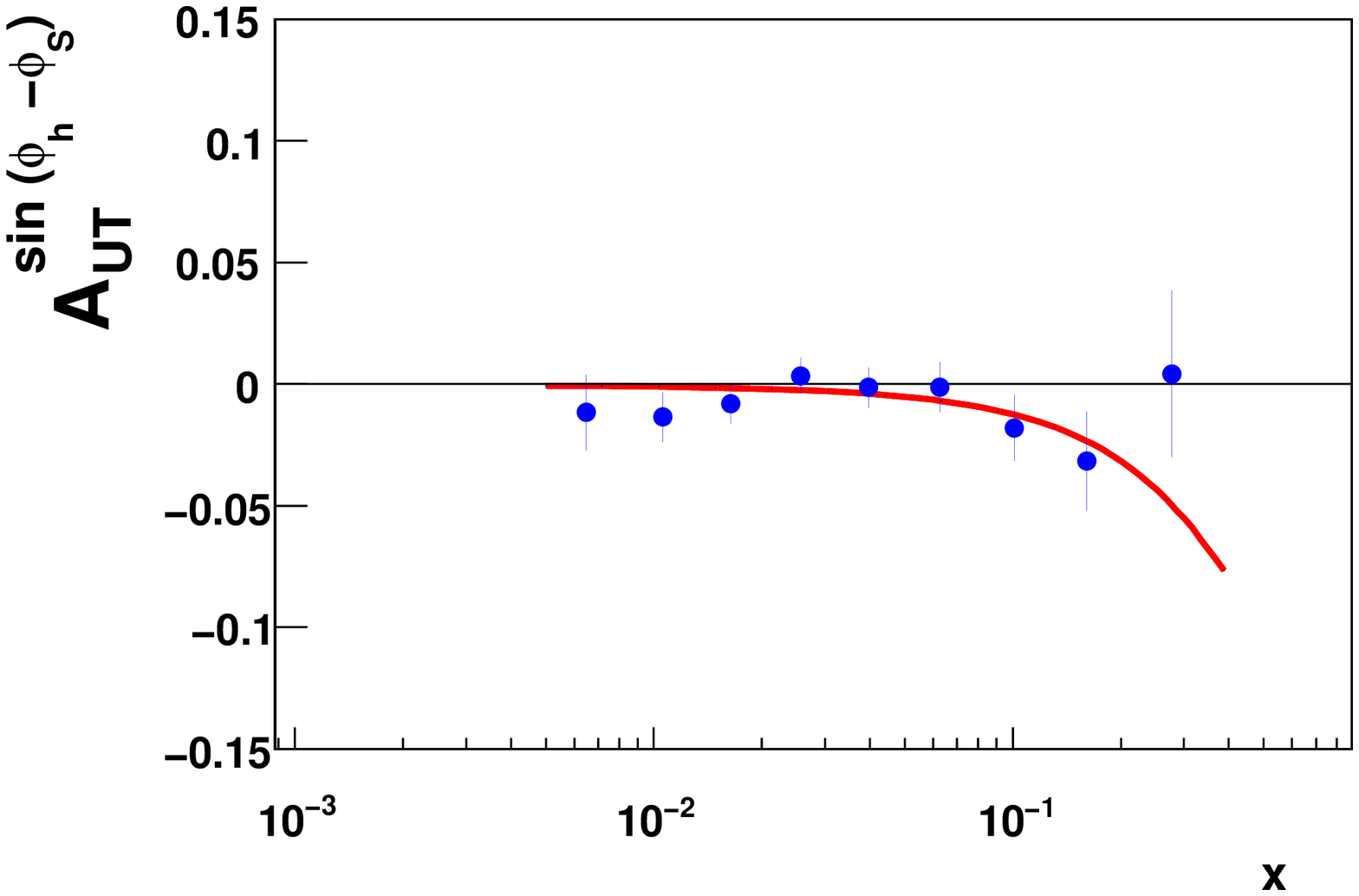}
  \\
  (a) & (b)
  \end{tabular}
\caption{Description of (a) HERMES \cite{:2009ti} and (b) COMPASS \cite{:2008dn} data on $\pi^+$ production as a function of $x_B$.}
\label{fig:sidis}
\end{figure*}

Let us now turn to the actual description of the experimental data. Our rather large $\chi^2$ with $\chi^2/{\rm d.o.f.} = 3.6$ has already indicated that the overall quality of our fitting is quite poor. In Fig.~\ref{fig:sidis} the result of the fit compared to $\pi^+$ HERMES \cite{:2009ti} and COMPASS \cite{:2008dn} data as a function of $x_B$. For $\pi^-$ asymmetry and $z_h$ and $P_{h\perp}$ dependences the description is comparable to that of Fig.~\ref{fig:sidis}. In other words, our description of SIDIS data is satisfactory, $\chi^2/\# data \sim 1.5$.

\begin{figure*}[hbt]
\centering
  \begin{tabular}{c@{\hspace*{5mm}}c}
    \includegraphics[scale=0.4]{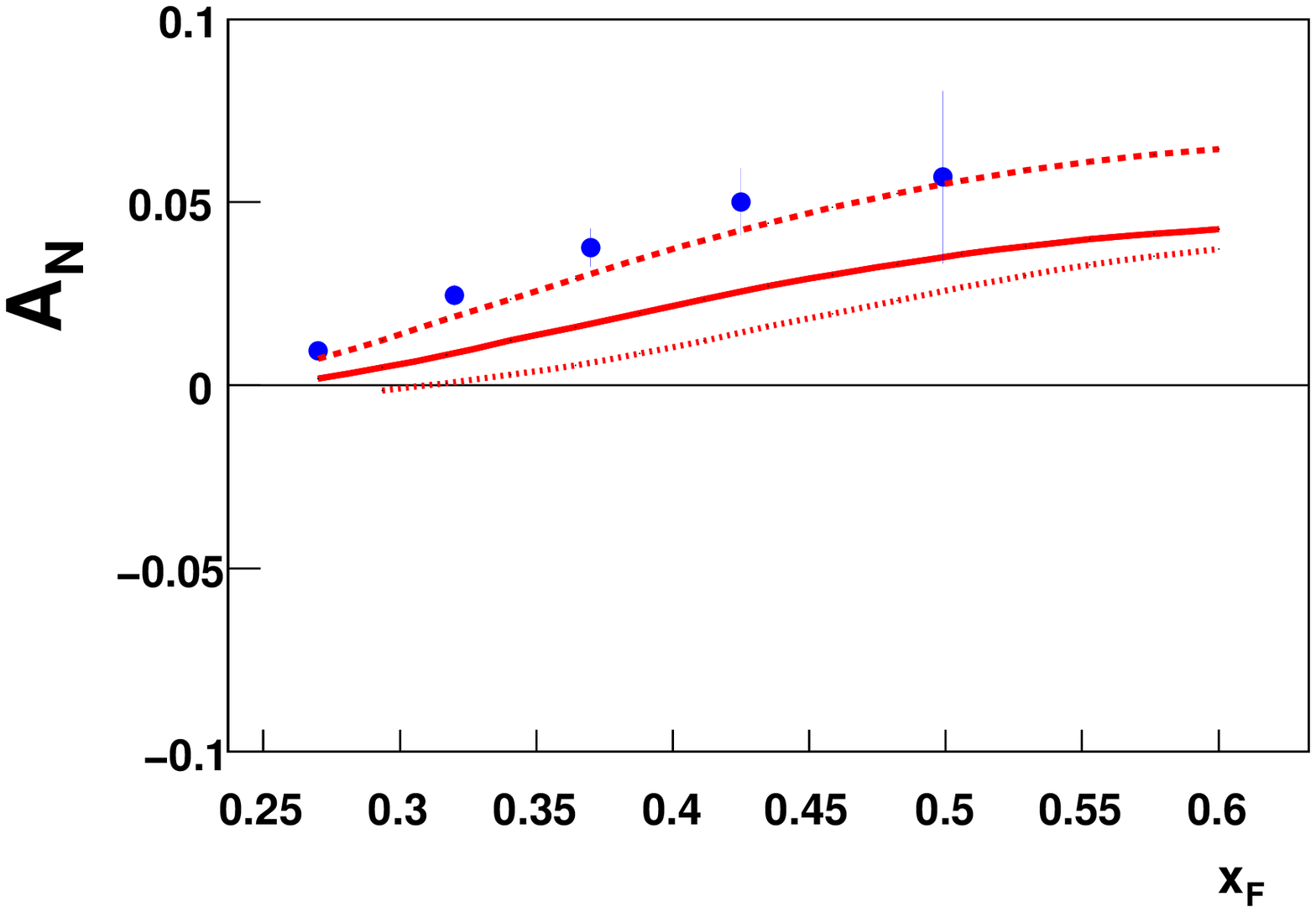}
    &
    \includegraphics[scale=0.4]{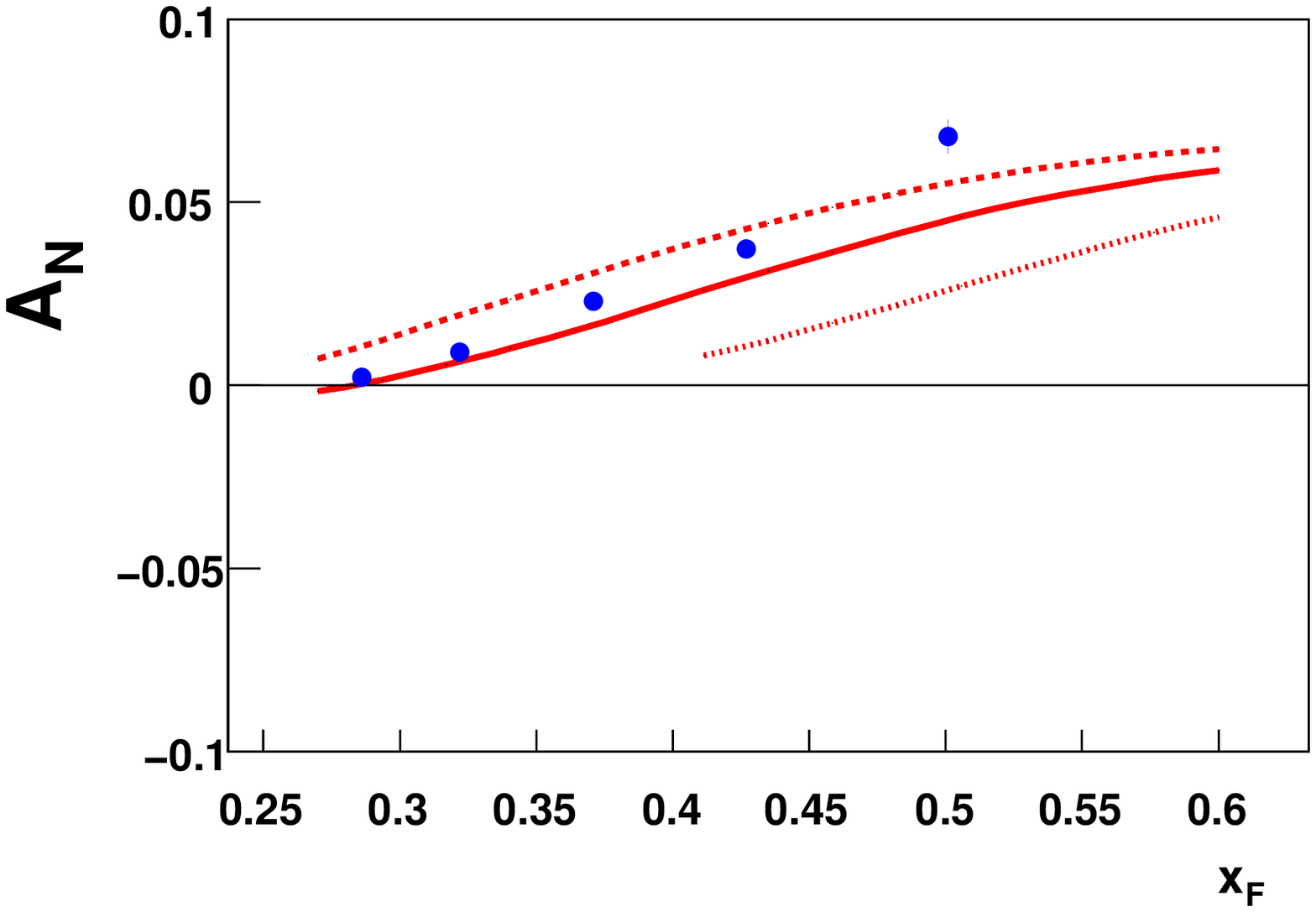}
  \\
      (a) & (b)
  \end{tabular}
\caption{Description of STAR $\pi^0$ data \cite{:2008qb} at rapidity (a) $y=3.3$ and (b) $y=3.7$ at $\sqrt{S}=200$ GeV.
Solid curves correspond to the scale $\mu=P_{h\perp}$, while dashed and dotted ones correspond to $\mu=P_{h\perp}/2$ and $\mu=2 P_{h\perp}$, respectively.}
\label{fig:pp1}
\end{figure*}

In Fig.~\ref{fig:pp1}, we compare the fit with the STAR $\pi^0$ data as a function of $x_F$ for $\sqrt{S}=200$ GeV at $y=3.3$ (a) and $y=3.7$ (b), respectively. The solid curves correspond to the scale $\mu=P_{h\perp}$. The description is reasonably good, though slightly worse than those for SIDIS data. We have also explored the theoretical uncertainty coming from the scale $\mu$ through its variation by a factor 2 up and down relative to the default values, and they are plotted as dashed and dotted curves in Fig.~\ref{fig:pp1}. This uncertainty is indeed quite large as one might expect since we are using the leading order formalism. The improvement could be achieved once the next-to-leading order calculations are performed~\cite{Vogelsang:2009pj,Kang:2008ey,Zhou:2008mz,Braun:2009mi,
Kang:2010xv}.
%%%%%%%%%%%%%%%%%%%%%%%%%%%%%%%%%%%
\begin{figure*}[hbt]
\centering
  \begin{tabular}{c@{\hspace*{5mm}}c}
    \includegraphics[scale=0.4]{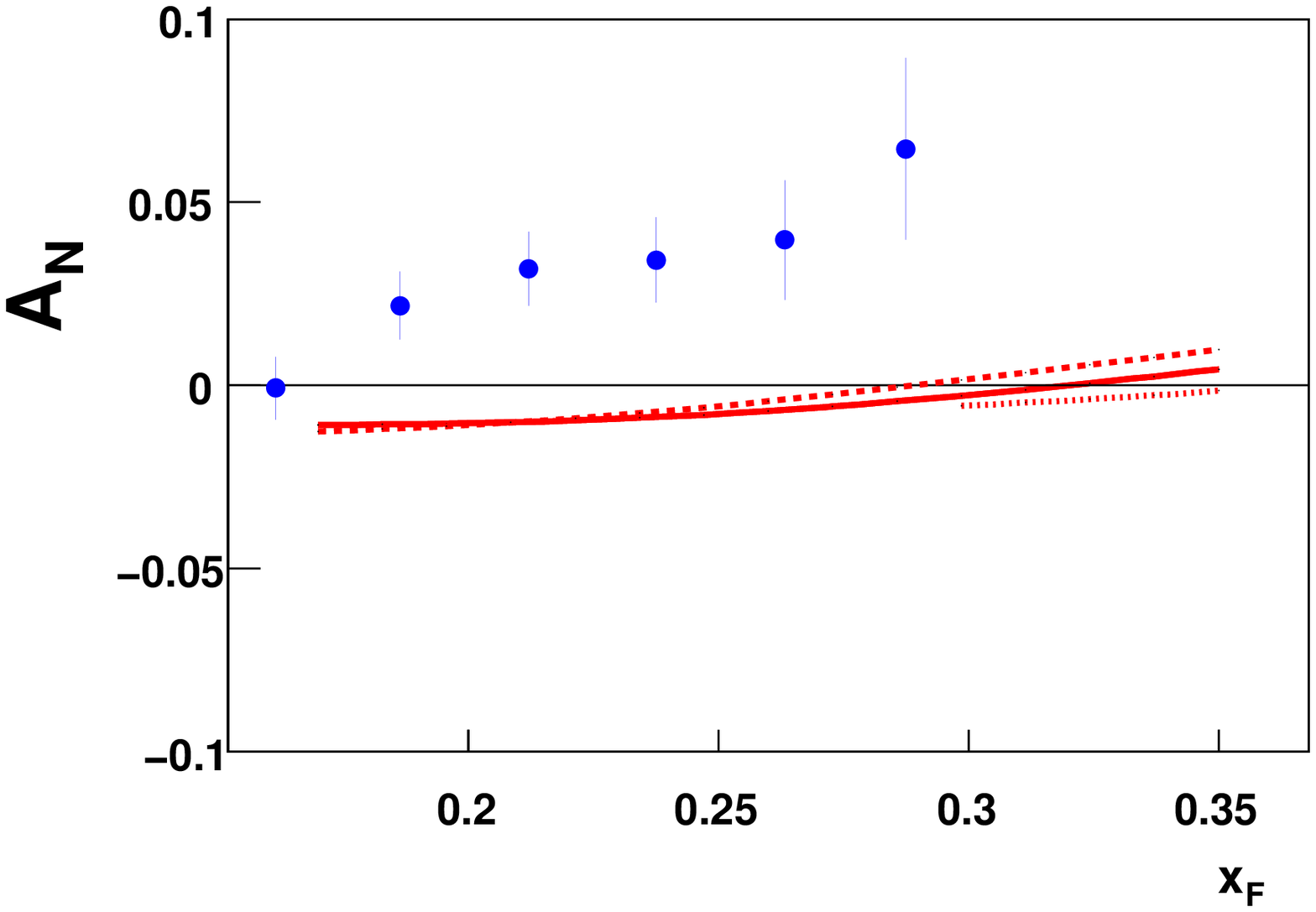}
    &
    \includegraphics[scale=0.4]{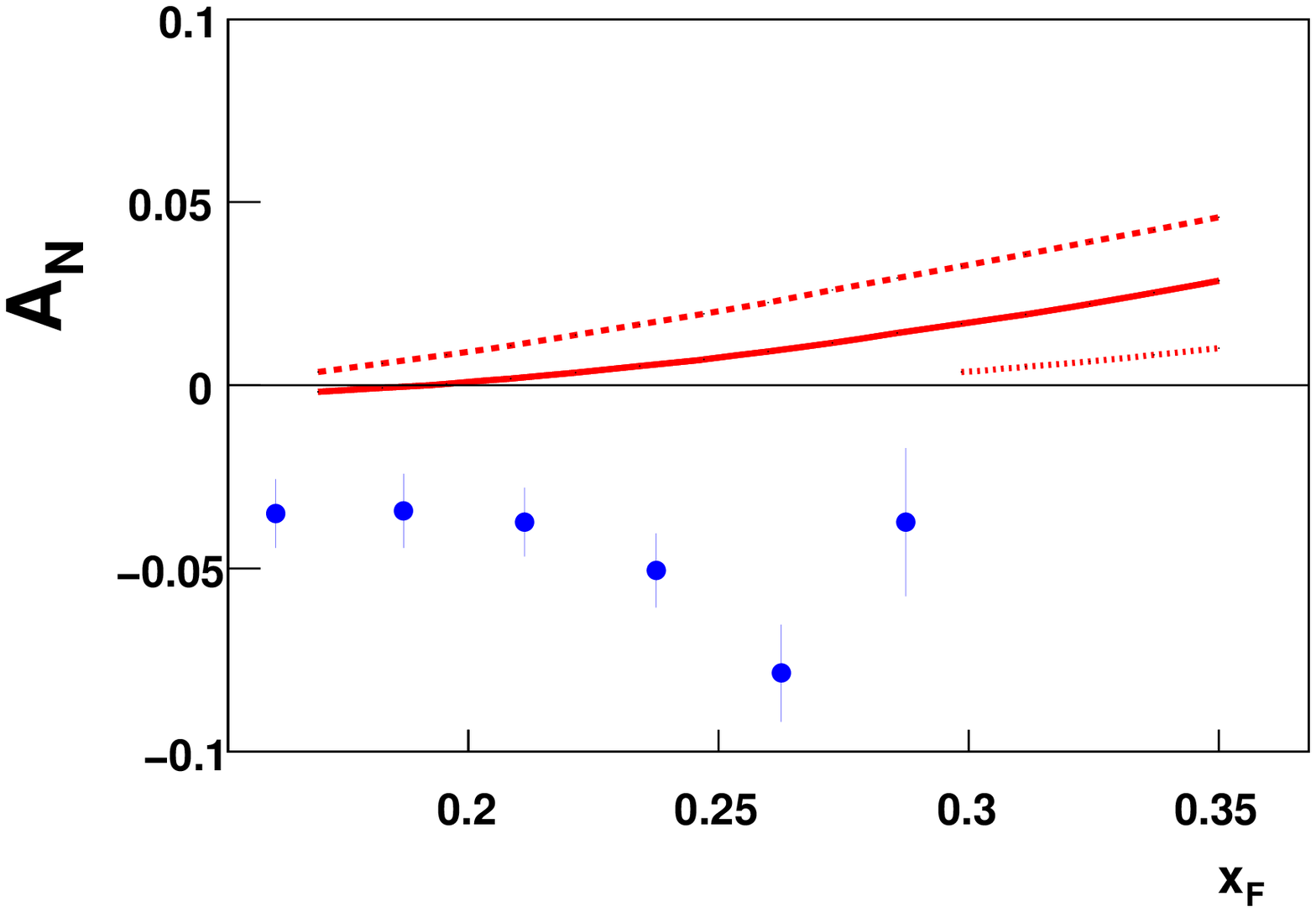}
  \\
      (a) & (b)
  \end{tabular}
\caption{Description of BRAHMS $\pi^+$ (a) and $\pi^-$ (b) data \cite{Lee:2007zzh} at forward angle $\theta=4^\circ$ at 
$\sqrt{S}=200$ GeV. Solid curves correspond to the scale $\mu=P_{h\perp}$, and dashed and dotted ones correspond to $\mu=P_{h\perp}/2$ and $\mu=2 P_{h\perp}$, respectively.}
\label{fig:pp2}
\end{figure*}

In Fig.~\ref{fig:pp2}, we compare the fit with the BRAHMS $\pi^+$ and  $\pi^-$ data at forward angle 
$\theta=4^{\circ}$ at $\sqrt{S}=200$ GeV. It is clear from these figures that our fitted parametrization for the Sivers functions (or the ETQS functions) is not compatible with the BRAHMS $\pi^+$ and  $\pi^-$ data, and even the signs for $A_N$ are opposite. It is worth pointing out that the previous measurements for the charged pion production in $pp$ collisions (e.g., those from E704 \cite{Adams:1991cs,Krueger:1998hz}) do have the consistent signs with BRAHMS. This finding is consistent with the heart of the ``sign mismatch'' paper~\cite{Kang:2011hk}. Our starting point for the possibility of node in $x$ is based on the fact that SIDIS and STAR data probe slightly different $x$ region: $x = x_B < 0.3$ for SIDIS, while $x\gtrsim 0.3$ in the integration for $x_F\gtrsim 0.3$ for STAR data. Thus a node in $x$ can describe both SIDIS and STAR data rather well. However, the BRAHMS has a $x$ region $x\sim x_F\in[0.15, 0.3]$ which overlaps with the SIDIS data. Thus the node in $x$ can not be a solution for the ``sign mismatch'' problem. Our failed attempt of the global fitting of both SIDIS and proton-proton data has once more confirmed that these data are not compatible with each other, if we consider only the twist-three contributions from the polarized nucleon. It indicates that there should be a sizable and more important contribution from the twist-three fragmentation in the produced hadrons \cite{Kang:2010zzb}. 

\begin{figure*}[hbt]
\centering
    \includegraphics[scale=0.4]{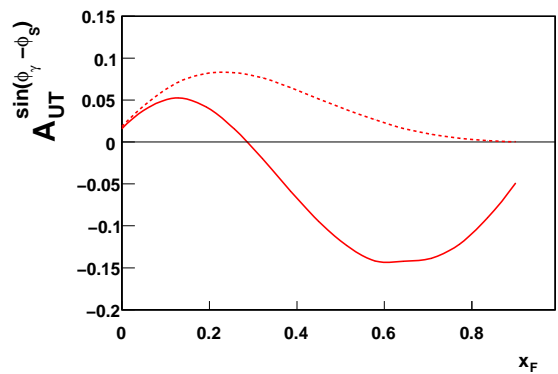}
\caption{Prediction of Drell-Yan asymmetry for RHIC kinematics $p^\uparrow p \to \ell^+\ell^-X$, $0<y<3$. Solid line corresponds to Sivers function with a node from this work and 
dashed line to Sivers function without node from Ref.~\cite{Anselmino:2008sga}. The same convention for the hadronic frame and asymmetry is used as 
in Ref.~\cite{Anselmino:2009st}.}
\label{fig:drellyan}
\end{figure*}

Even though we fail to cure the ``sign mismatch'' problem from the node in $x$ scenario, the concept that the Sivers function does not need to have the same sign in the whole kinematic region (either $x$ or $k_\perp$) has important implications, especially when it comes to check experimentally the sign change of the Sivers function from SIDIS to DY processes. In Fig.~\ref{fig:drellyan}, we show the calculation of DY asymmetry for RHIC kinematics at $\sqrt{S}=200$ GeV as a function of $x_F$. The solid curve corresponds to the calculations using the Sivers function with a node in $x$ from Table.~\ref{table:I}, and the dashed curve is the calculation based on the Sivers function from \cite{Anselmino:2008sga} which has no node in $x$. One can see that the prediction changes drastically in case node is present, however in the region of $0<x_F<0.25$ the sign of the asymmetry is consistent and dictated by the Sivers function constrained from SIDIS measurements. Regardless of possible nodes
this region is safe for measurement. In the future DY experiments, the $Q^2$ range will also be different, to have a solid prediction, one of course also needs to include the effect of the evolution \cite{Kang:2011mr,Aybat:2011ge,Aybat:2011ta}.

%%%%%%%%%%%%%%%%%%%%%%%%%%%%%%%%%%% 
\subsection{Exploration of node in $k_\perp$: the simplest study}
\begin{figure*}[hbt]
\centering
    \includegraphics[scale=0.5]{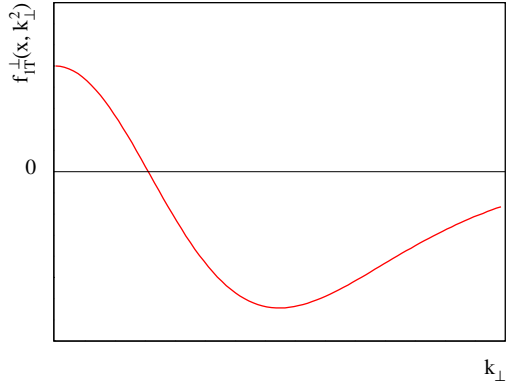}
\caption{An illustration of the Sivers function with a node in $k_\perp$.}
\label{fig:kt}
\end{figure*}

The Sivers function with a node in $k_\perp$ has also been suggested as a solution to the ``sign mismatch'' problem in Ref.~\cite{Kang:2011hk}. The main idea comes from the fact that HERMES and COMPASS SIDIS data are mostly relevant for the extraction of the Sivers function $f^{\perp q}_{1T}(x,k_\perp^2)$ at relatively modest $Q^2\sim 2.5 - 3.5$ GeV$^2$. Since the TMD factorization formalism is valid only for $k_\perp\ll Q$, the data thus constrain the function and its sign only at a very low $k_\perp\sim \Lambda_{\rm QCD}$. However, to obtain a functional form for ETQS function $T_{q,F}(x,x)$, one needs to integrate over the full range of $k_\perp$ in Eq.~\eqref{eq:relation}. Since currently we have assumed a Gaussian form for the $k_\perp$-dependence which has the same sign for the whole $k_\perp$ region, the $k_\perp$-integration will have the same sign of low $k_\perp$ part. 
However, if somehow the high $k_\perp$ region has opposite sign to the low $k_\perp$ part, this might alter the sign of the $k_\perp$-moment in the integration and thus lead to the correct sign of $T_{q,F}(x,x)$. See Fig.~\ref{fig:kt} for an illustration.

In this subsection we explore such a possibility. Talking about the $k_\perp$-dependence, it is important to recall again that the relation in Eq.~\eqref{eq:relation} is subject to the ultraviolet (UV) subtraction and the adopted factorization scheme. To avoid such a problem, as a natural extension to the usual Gaussian form, we choose $k_\perp$-dependence as a difference between two Gaussian functions with slightly different widths. This is the simplest case which allows a node in $k_\perp$, and we will explore whether this simple extension works in practice. The Sivers function will now be parametrized as follows:
\ben
f_{1T}^{\perp q}(x, k_\perp^2) &=& - \, {\cal N}_q(x)  h(k_\perp) f_{q/A} (x, k_\perp^2),
\een
where we use the usual $x$-dependence (without node) for simplicity
\ben
{\cal N}_q(x) =  N_q x^{\alpha_q}(1-x)^{\beta_q}
\frac{(\alpha_q+\beta_q)^{(\alpha_q+\beta_q)}}
{\alpha_q^{\alpha_q} \beta_q^{\beta_q}}.
\een
However, the $k_\perp$-dependence $h(k_\perp)$ will be changed to (from Eq.~\eqref{eq:siverskt})
\ben
h(k_\perp) = \sqrt{2e}\,M\left[\frac{e^{-\mathbf{k}_\perp^2/{M_{1}^2}}}{M_1}-\frac{e^{-\mathbf{k}_\perp^2/{M_{2}^2}}}{M_2}\right].
\een
One has to choose $M_2>M_1$, thus the low-$k_\perp$ will be positive, i.e., follow the same sign like the usual Sivers function as in Eq.~\eqref{eq:sivfac}, and the $k_\perp$-shape follows Fig.~\ref{fig:kt}. 

\begin{figure*}[hbt]
\centering
    \includegraphics[scale=0.8]{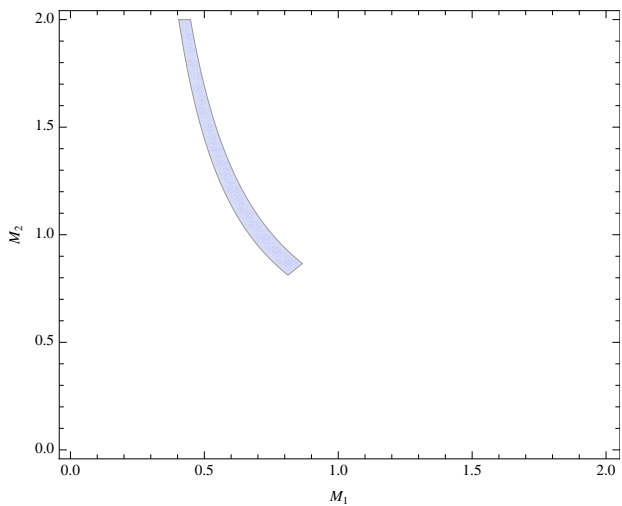}
\caption{The allowed region for parameters $M_1$ and $M_2$ at $P_{h\perp}=0.5$ GeV and $z_h=0.5$.}
\label{fig:region}
\end{figure*}

Using Eq.~\eqref{eq:relation}, we could derive the functional form for the ETQS function, we have
\ben
T_{q,F}(x, x)=\sqrt{2e}\la k_\perp^2\ra \left[\frac{M_1^3}{\left(\la k_\perp^2\ra + M_1^2\right)^2} - 
\frac{M_2^3}{\left(\la k_\perp^2\ra + M_2^2\right)^2} \right]{\cal N}_q(x) f_{q/A} (x).
\een
In order that the sign of $T_{q,F}(x, x)$ is altered, we should have
\ben
 \frac{M_2^3}{\left(\la k_\perp^2\ra + M_2^2\right)^2} > \frac{M_1^3}{\left(\la k_\perp^2\ra + M_1^2\right)^2}.
 \label{eq:req1}
\een
We could also derive the expression for the Sivers asymmetry in SIDIS process, and find
\ben
F_{UT,T}^{\sin(\phi_h -\phi_s)}\;
 &=& x_B z_h P_{h\perp} 
\sum_q e_q^2\, {\cal N}_q(x_B) f_{q/A}(x_B) D_{h/q}(z_h) \frac{\sqrt{2e}\la k_\perp^2\ra}{\pi} 
\nonumber\\
&&\times
\left[
\frac{M_1^3}{\left(\la k_\perp^2\ra + M_1^2\right)^2} \frac{1}{\la P_{h\perp1}^2\ra^2} e^{-\frac{P_{h\perp}^2}{\la P_{h\perp1}^2\ra}}-\frac{M_2^3}{\left(\la k_\perp^2\ra + M_2^2\right)^2} \frac{1}{\la P_{h\perp2}^2\ra^2} e^{-\frac{P_{h\perp}^2}{\la P_{h\perp2}^2\ra}}\right],
\een
where the widths $\la P_{h\perp i}^2\ra$ ($i=1,2$) is defined as
\ben
\la P_{h\perp i}^2\ra = \la p_T^2\ra + z_h^2 \frac{\la k_\perp^2\ra M_i^2}{\la k_\perp^2\ra + M_i^2}.
\een
In order that the asymmetry follows the same sign at low $P_{h\perp}$ like before, one requires
\ben
 \frac{M_2^3}{\left(M_2^2+\gamma(z)\la k_\perp^2\ra \right)^2} < \frac{M_1^3}{\left(M_1^2+ \gamma(z)\la k_\perp^2\ra \right)^2},
\label{eq:req2}
\een
where $\gamma(z)^{-1}=1+z_h^2\la k_\perp^2\ra/\la p_T^2\ra$. Thus we have three requirements from Eqs.~\eqref{eq:req1}, \eqref{eq:req2} plus $M_2>M_1$. One also need to take into account the fact that the Sivers asymmetries measured by both HERMES and COMPASS do not change sign up to $P_{h\perp}\sim 1$ GeV. All these requirements have constrained the allowed parameter space for $M_1$ and $M_2$ to a very limited (small) region. For an illustration, see Fig.~\ref{fig:region} for a typical $P_{h\perp}=0.5$ GeV and $z_h=0.5$. This region gets even smaller if $P_{h\perp}$ increases and/or $z_h$ decreases. From such a simple study, we find that our simplest extension to allow a node in $k_\perp$ seems {\it not} to be a natural solution to the ``sign puzzle''. Of course, other types of $k_\perp$ dependence which also has node in $k_\perp$ might still be possible\footnote{We have also explored another $k_\perp$-dependent form \cite{boer}: $h(k_\perp)=\sqrt{2e}\frac{M}{M_1}e^{-\mathbf{k}_\perp^2/{M_{1}^2}}(1-\eta\, {\mathbf k}_\perp^2)$ and we find that the allowed parameter space for $M_1$ and $\eta$ is again very small.}. 
At the end, we want to emphasize again that there is the important UV regularization issue, which is out of the scope of our current study.

To finish this section, we study the possibility of a node in $x$ or a node in $k_\perp$ in the Sivers function. The simplest extensions to contain such a node seem not to be working for both cases. This strongly suggests that there could be a sizable contribution from the twist-three fragmentation in the single inclusive hadron production \cite{Kang:2010zzb}, and we hope future experimental study could give us clear answers.

\section{Conclusions}
In this paper, for the first time, we make an attempt for a global fitting of both SIDIS and proton-proton data on the spin asymmetries. We use a TMD factorization formalism to describe the SIDIS Sivers asymmetry for the hadron production at low $P_{h\perp}$, and the collinear twist-three factorization formalism for the proton-proton data on the single inclusive hadron production at high $P_{h\perp}$. 
We adopt a more flexible functional form for the Sivers function in order to simultaneously describe both SIDIS and proton-proton data. By including only the contribution from the so-called soft gluon pole ETQS function in the polarized nucleon, we find that we are not able to describe well all the data. While all SIDIS data and STAR data on $\pi^0$ production can be explained, node in $x$ does not account for BRAHMS proton-proton data for the $\pi^+$ and $\pi^-$ production.
We have explored the possibility of a node in $x$ or a node in $k_\perp$ for the Sivers function. Our simplest extensions to allow a node for both cases seem not able to cure the ``sign mismatch'' problem.
If leaving behind the UV regularization issue for the relation between TMDs and collinear functions, we will conclude that there could be a sizable contribution from the twist-three fragmentation function for the single inclusive hadron production in proton-proton collisions. We hope the future experiments could give us clear answers.

A side-effect (lesson) learned from our study is that one should be very careful in extrapolating the Sivers function (or any other TMDs which do not need to be positive) to the region where there is no experimental measurements or constrains. The size and the sign of the functions in these region should be carefully measured in the future experiments. For example, careful analysis of SIDIS data at large values of $x_B$ is needed in order to rule out or confirm the possible node of the Sivers function. Also measurement of 
$\pi^\pm$ $A_N$ at larger values of $x_F$ is needed in order to confirm that the node (in $x$) is not compatible with BRAHMS data. Careful analysis of $k_\perp$ dependence of TMDs is also needed.

\section*{Acknowledgement}
We thank Daniel Boer, Andreas Metz, Jian-Wei Qiu, Werner Vogelsang, Feng Yuan and Ted Rogers for discussions and careful reading of the manuscript.
This work was supported in part by the U.S. Department of Energy under Contract No.~DE-AC02-98CH10886 (ZK) and DE-AC05-06OR23177 (AP).

\bibliography{biblio}

%merlin.mbs 2010-03-15 4.21a (PWD, AO, DPC)
%Control: key (0)
%Control: author (8) initials jnrlst
%Control: editor formatted (1) identically to author
%Control: production of article title (-1) disabled
%Control: page (0) single
%Control: year (1) truncated
%Control: production of eprint (0) enabled
\begin{thebibliography}{65}%
\makeatletter
\providecommand \@ifxundefined [1]{%
 \@ifx{#1\undefined}
}%
\providecommand \@ifnum [1]{%
 \ifnum #1\expandafter \@firstoftwo
 \else \expandafter \@secondoftwo
 \fi
}%
\providecommand \@ifx [1]{%
 \ifx #1\expandafter \@firstoftwo
 \else \expandafter \@secondoftwo
 \fi
}%
\providecommand \natexlab [1]{#1}%
\providecommand \enquote  [1]{``#1''}%
\providecommand \bibnamefont  [1]{#1}%
\providecommand \bibfnamefont [1]{#1}%
\providecommand \citenamefont [1]{#1}%
\providecommand \href@noop [0]{\@secondoftwo}%
\providecommand \href [0]{\begingroup \@sanitize@url \@href}%
\providecommand \@href[1]{\@@startlink{#1}\@@href}%
\providecommand \@@href[1]{\endgroup#1\@@endlink}%
\providecommand \@sanitize@url [0]{\catcode `\\12\catcode `\$12\catcode
  `\&12\catcode `\#12\catcode `\^12\catcode `\_12\catcode `\%12\relax}%
\providecommand \@@startlink[1]{}%
\providecommand \@@endlink[0]{}%
\providecommand \url  [0]{\begingroup\@sanitize@url \@url }%
\providecommand \@url [1]{\endgroup\@href {#1}{\urlprefix }}%
\providecommand \urlprefix  [0]{URL }%
\providecommand \Eprint [0]{\href }%
\@ifxundefined \urlstyle {%
  \providecommand \doi  [0]{\begingroup \@sanitize@url \@doi}%
  \providecommand \@doi [1]{\endgroup \@@startlink {\doibase
  #1}doi:\discretionary {}{}{}#1\@@endlink }%
}{%
  \providecommand \doi  [0]{doi:\discretionary{}{}{}\begingroup
  \urlstyle{rm}\Url }%
}%
\providecommand \doibase [0]{http://dx.doi.org/}%
\providecommand \Doi [0]{\begingroup \@sanitize@url \@Doi }%
\providecommand \@Doi  [1]{\endgroup\@@startlink{\doibase#1}\@@Doi}%
\providecommand \@@Doi [1]{#1\@@endlink}%
\providecommand \selectlanguage [0]{\@gobble}%
\providecommand \bibinfo  [0]{\@secondoftwo}%
\providecommand \bibfield  [0]{\@secondoftwo}%
\providecommand \translation [1]{[#1]}%
\providecommand \BibitemOpen [0]{}%
\providecommand \bibitemStop [0]{}%
\providecommand \bibitemNoStop [0]{.\EOS\space}%
\providecommand \EOS [0]{\spacefactor3000\relax}%
\providecommand \BibitemShut  [1]{\csname bibitem#1\endcsname}%
%</preamble>
\bibitem [{\citenamefont {Boer}\ \emph {et~al.}(2011)\citenamefont {Boer} \emph
  {et~al.}}]{Boer:2011fh}%
  \BibitemOpen
  \bibfield  {author} {\bibinfo {author} {\bibfnamefont {D.}~\bibnamefont
  {Boer}} \emph {et~al.},\ }\href@noop {} { (\bibinfo {year} {2011})},\ \Eprint
  {http://arxiv.org/abs/1108.1713} {arXiv:1108.1713 [nucl-th]} \BibitemShut
  {NoStop}%
%%CITATION = 1108.1713;%%
\bibitem [{\citenamefont {Efremov}\ and\ \citenamefont
  {Teryaev}(1982)}]{Efremov:1981sh}%
  \BibitemOpen
  \bibfield  {author} {\bibinfo {author} {\bibfnamefont {A.~V.}\ \bibnamefont
  {Efremov}}\ and\ \bibinfo {author} {\bibfnamefont {O.~V.}\ \bibnamefont
  {Teryaev}},\ }\href@noop {} {\bibfield  {journal} {\bibinfo  {journal} {Sov.
  J. Nucl. Phys.},\ }\textbf {\bibinfo {volume} {36}},\ \bibinfo {pages} {140}
  (\bibinfo {year} {1982})}\BibitemShut {NoStop}%
%%CITATION = SJNCA,36,140;%%
\bibitem [{\citenamefont {Efremov}\ and\ \citenamefont
  {Teryaev}(1985)}]{Efremov:1984ip}%
  \BibitemOpen
  \bibfield  {author} {\bibinfo {author} {\bibfnamefont {A.}~\bibnamefont
  {Efremov}}\ and\ \bibinfo {author} {\bibfnamefont {O.}~\bibnamefont
  {Teryaev}},\ }\Doi {10.1016/0370-2693(85)90999-2} {\bibfield  {journal}
  {\bibinfo  {journal} {Phys.Lett.},\ }\textbf {\bibinfo {volume} {B150}},\
  \bibinfo {pages} {383} (\bibinfo {year} {1985})}\BibitemShut {NoStop}%
%%CITATION = PHLTA,B150,383;%%
\bibitem [{\citenamefont {Qiu}\ and\ \citenamefont
  {Sterman}(1991)}]{Qiu:1991pp}%
  \BibitemOpen
  \bibfield  {author} {\bibinfo {author} {\bibfnamefont {J.-w.}\ \bibnamefont
  {Qiu}}\ and\ \bibinfo {author} {\bibfnamefont {G.}~\bibnamefont {Sterman}},\
  }\Doi {10.1103/PhysRevLett.67.2264} {\bibfield  {journal} {\bibinfo
  {journal} {Phys. Rev. Lett.},\ }\textbf {\bibinfo {volume} {67}},\ \bibinfo
  {pages} {2264} (\bibinfo {year} {1991})}\BibitemShut {NoStop}%
%%CITATION = PRLTA,67,2264;%%
\bibitem [{\citenamefont {Qiu}\ and\ \citenamefont
  {Sterman}(1999)}]{Qiu:1998ia}%
  \BibitemOpen
  \bibfield  {author} {\bibinfo {author} {\bibfnamefont {J.-w.}\ \bibnamefont
  {Qiu}}\ and\ \bibinfo {author} {\bibfnamefont {G.~F.}\ \bibnamefont
  {Sterman}},\ }\Doi {10.1103/PhysRevD.59.014004} {\bibfield  {journal}
  {\bibinfo  {journal} {Phys.Rev.},\ }\textbf {\bibinfo {volume} {D59}},\
  \bibinfo {pages} {014004} (\bibinfo {year} {1999})},\ \Eprint
  {http://arxiv.org/abs/hep-ph/9806356} {arXiv:hep-ph/9806356 [hep-ph]}
  \BibitemShut {NoStop}%
%%CITATION = HEP-PH/9806356;%%
\bibitem [{\citenamefont {Koike}\ and\ \citenamefont
  {Tomita}(2009)}]{Koike:2009ge}%
  \BibitemOpen
  \bibfield  {author} {\bibinfo {author} {\bibfnamefont {Y.}~\bibnamefont
  {Koike}}\ and\ \bibinfo {author} {\bibfnamefont {T.}~\bibnamefont {Tomita}},\
  }\Doi {10.1016/j.physletb.2009.04.017} {\bibfield  {journal} {\bibinfo
  {journal} {Phys.Lett.},\ }\textbf {\bibinfo {volume} {B675}},\ \bibinfo
  {pages} {181} (\bibinfo {year} {2009})},\ \Eprint
  {http://arxiv.org/abs/0903.1923} {arXiv:0903.1923 [hep-ph]} \BibitemShut
  {NoStop}%
%%CITATION = ARXIV:0903.1923;%%
\bibitem [{\citenamefont {Kang}\ \emph {et~al.}(2010)\citenamefont {Kang},
  \citenamefont {Yuan},\ and\ \citenamefont {Zhou}}]{Kang:2010zzb}%
  \BibitemOpen
  \bibfield  {author} {\bibinfo {author} {\bibfnamefont {Z.-B.}\ \bibnamefont
  {Kang}}, \bibinfo {author} {\bibfnamefont {F.}~\bibnamefont {Yuan}}, \ and\
  \bibinfo {author} {\bibfnamefont {J.}~\bibnamefont {Zhou}},\ }\Doi
  {10.1016/j.physletb.2010.07.003} {\bibfield  {journal} {\bibinfo  {journal}
  {Phys. Lett.},\ }\textbf {\bibinfo {volume} {B691}},\ \bibinfo {pages} {243}
  (\bibinfo {year} {2010})},\ \Eprint {http://arxiv.org/abs/1002.0399}
  {arXiv:1002.0399 [hep-ph]} \BibitemShut {NoStop}%
%%CITATION = 1002.0399;%%
\bibitem [{\citenamefont {Kouvaris}\ \emph {et~al.}(2006)\citenamefont
  {Kouvaris}, \citenamefont {Qiu}, \citenamefont {Vogelsang},\ and\
  \citenamefont {Yuan}}]{Kouvaris:2006zy}%
  \BibitemOpen
  \bibfield  {author} {\bibinfo {author} {\bibfnamefont {C.}~\bibnamefont
  {Kouvaris}}, \bibinfo {author} {\bibfnamefont {J.-W.}\ \bibnamefont {Qiu}},
  \bibinfo {author} {\bibfnamefont {W.}~\bibnamefont {Vogelsang}}, \ and\
  \bibinfo {author} {\bibfnamefont {F.}~\bibnamefont {Yuan}},\ }\Doi
  {10.1103/PhysRevD.74.114013} {\bibfield  {journal} {\bibinfo  {journal}
  {Phys. Rev.},\ }\textbf {\bibinfo {volume} {D74}},\ \bibinfo {pages} {114013}
  (\bibinfo {year} {2006})},\ \Eprint {http://arxiv.org/abs/hep-ph/0609238}
  {arXiv:hep-ph/0609238} \BibitemShut {NoStop}%
%%CITATION = HEP-PH/0609238;%%
\bibitem [{\citenamefont {Kanazawa}\ and\ \citenamefont
  {Koike}(2011)}]{Kanazawa:2011bg}%
  \BibitemOpen
  \bibfield  {author} {\bibinfo {author} {\bibfnamefont {K.}~\bibnamefont
  {Kanazawa}}\ and\ \bibinfo {author} {\bibfnamefont {Y.}~\bibnamefont
  {Koike}},\ }\Doi {10.1103/PhysRevD.83.114024} {\bibfield  {journal} {\bibinfo
   {journal} {Phys. Rev.},\ }\textbf {\bibinfo {volume} {D83}},\ \bibinfo
  {pages} {114024} (\bibinfo {year} {2011})},\ \Eprint
  {http://arxiv.org/abs/1104.0117} {arXiv:1104.0117 [hep-ph]} \BibitemShut
  {NoStop}%
%%CITATION = 1104.0117;%%
\bibitem [{\citenamefont {Ji}\ \emph {et~al.}(2005)\citenamefont {Ji},
  \citenamefont {Ma},\ and\ \citenamefont {Yuan}}]{Ji:2004wu}%
  \BibitemOpen
  \bibfield  {author} {\bibinfo {author} {\bibfnamefont {X.-d.}\ \bibnamefont
  {Ji}}, \bibinfo {author} {\bibfnamefont {J.-p.}\ \bibnamefont {Ma}}, \ and\
  \bibinfo {author} {\bibfnamefont {F.}~\bibnamefont {Yuan}},\ }\href@noop {}
  {\bibfield  {journal} {\bibinfo  {journal} {Phys. Rev.},\ }\textbf {\bibinfo
  {volume} {D71}},\ \bibinfo {pages} {034005} (\bibinfo {year} {2005})},\
  \Eprint {http://arxiv.org/abs/hep-ph/0404183} {hep-ph/0404183} \BibitemShut
  {NoStop}%
%%CITATION = HEP-PH 0404183;%%
\bibitem [{\citenamefont {Ji}\ \emph {et~al.}(2004)\citenamefont {Ji},
  \citenamefont {Ma},\ and\ \citenamefont {Yuan}}]{Ji:2004xq}%
  \BibitemOpen
  \bibfield  {author} {\bibinfo {author} {\bibfnamefont {X.-d.}\ \bibnamefont
  {Ji}}, \bibinfo {author} {\bibfnamefont {J.-P.}\ \bibnamefont {Ma}}, \ and\
  \bibinfo {author} {\bibfnamefont {F.}~\bibnamefont {Yuan}},\ }\Doi
  {10.1016/j.physletb.2004.07.026} {\bibfield  {journal} {\bibinfo  {journal}
  {Phys. Lett.},\ }\textbf {\bibinfo {volume} {B597}},\ \bibinfo {pages} {299}
  (\bibinfo {year} {2004})},\ \Eprint {http://arxiv.org/abs/hep-ph/0405085}
  {arXiv:hep-ph/0405085} \BibitemShut {NoStop}%
%%CITATION = HEP-PH/0405085;%%
\bibitem [{\citenamefont {Sivers}(1990)}]{Sivers:1989cc}%
  \BibitemOpen
  \bibfield  {author} {\bibinfo {author} {\bibfnamefont {D.~W.}\ \bibnamefont
  {Sivers}},\ }\href@noop {} {\bibfield  {journal} {\bibinfo  {journal} {Phys.
  Rev.},\ }\textbf {\bibinfo {volume} {D41}},\ \bibinfo {pages} {83} (\bibinfo
  {year} {1990})}\BibitemShut {NoStop}%
%%CITATION = PHRVA,D41,83;%%
\bibitem [{\citenamefont {Sivers}(1991)}]{Sivers:1990fh}%
  \BibitemOpen
  \bibfield  {author} {\bibinfo {author} {\bibfnamefont {D.~W.}\ \bibnamefont
  {Sivers}},\ }\href@noop {} {\bibfield  {journal} {\bibinfo  {journal} {Phys.
  Rev.},\ }\textbf {\bibinfo {volume} {D43}},\ \bibinfo {pages} {261} (\bibinfo
  {year} {1991})}\BibitemShut {NoStop}%
%%CITATION = PHRVA,D43,261;%%
\bibitem [{\citenamefont {Boer}\ and\ \citenamefont
  {Mulders}(1998)}]{Boer:1997nt}%
  \BibitemOpen
  \bibfield  {author} {\bibinfo {author} {\bibfnamefont {D.}~\bibnamefont
  {Boer}}\ and\ \bibinfo {author} {\bibfnamefont {P.~J.}\ \bibnamefont
  {Mulders}},\ }\Doi {10.1103/PhysRevD.57.5780} {\bibfield  {journal} {\bibinfo
   {journal} {Phys. Rev.},\ }\textbf {\bibinfo {volume} {D57}},\ \bibinfo
  {pages} {5780} (\bibinfo {year} {1998})},\ \Eprint
  {http://arxiv.org/abs/hep-ph/9711485} {arXiv:hep-ph/9711485} \BibitemShut
  {NoStop}%
%%CITATION = HEP-PH/9711485;%%
\bibitem [{\citenamefont {Efremov}\ \emph {et~al.}(2005)\citenamefont
  {Efremov}, \citenamefont {Goeke}, \citenamefont {Menzel}, \citenamefont
  {Metz},\ and\ \citenamefont {Schweitzer}}]{Efremov:2004tp}%
  \BibitemOpen
  \bibfield  {author} {\bibinfo {author} {\bibfnamefont {A.~V.}\ \bibnamefont
  {Efremov}}, \bibinfo {author} {\bibfnamefont {K.}~\bibnamefont {Goeke}},
  \bibinfo {author} {\bibfnamefont {S.}~\bibnamefont {Menzel}}, \bibinfo
  {author} {\bibfnamefont {A.}~\bibnamefont {Metz}}, \ and\ \bibinfo {author}
  {\bibfnamefont {P.}~\bibnamefont {Schweitzer}},\ }\Doi
  {10.1016/j.physletb.2005.03.010} {\bibfield  {journal} {\bibinfo  {journal}
  {Phys. Lett.},\ }\textbf {\bibinfo {volume} {B612}},\ \bibinfo {pages} {233}
  (\bibinfo {year} {2005})},\ \Eprint {http://arxiv.org/abs/hep-ph/0412353}
  {arXiv:hep-ph/0412353} \BibitemShut {NoStop}%
%%CITATION = HEP-PH/0412353;%%
\bibitem [{\citenamefont {Vogelsang}\ and\ \citenamefont
  {Yuan}(2005)}]{Vogelsang:2005cs}%
  \BibitemOpen
  \bibfield  {author} {\bibinfo {author} {\bibfnamefont {W.}~\bibnamefont
  {Vogelsang}}\ and\ \bibinfo {author} {\bibfnamefont {F.}~\bibnamefont
  {Yuan}},\ }\href@noop {} {\bibfield  {journal} {\bibinfo  {journal} {Phys.
  Rev.},\ }\textbf {\bibinfo {volume} {D72}},\ \bibinfo {pages} {054028}
  (\bibinfo {year} {2005})}\BibitemShut {NoStop}%
%%CITATION = HEP-PH 0507266;%%
\bibitem [{\citenamefont {Anselmino}\ \emph {et~al.}(2005)\citenamefont
  {Anselmino} \emph {et~al.}}]{Anselmino:2005nn}%
  \BibitemOpen
  \bibfield  {author} {\bibinfo {author} {\bibfnamefont {M.}~\bibnamefont
  {Anselmino}} \emph {et~al.},\ }\Doi {10.1103/PhysRevD.71.074006} {\bibfield
  {journal} {\bibinfo  {journal} {Phys. Rev.},\ }\textbf {\bibinfo {volume}
  {D71}},\ \bibinfo {pages} {074006} (\bibinfo {year} {2005})},\ \Eprint
  {http://arxiv.org/abs/hep-ph/0501196} {arXiv:hep-ph/0501196} \BibitemShut
  {NoStop}%
%%CITATION = HEP-PH/0501196;%%
\bibitem [{\citenamefont {Arnold}\ \emph {et~al.}(2008)\citenamefont {Arnold},
  \citenamefont {Efremov}, \citenamefont {Goeke}, \citenamefont {Schlegel},\
  and\ \citenamefont {Schweitzer}}]{Arnold:2008ap}%
  \BibitemOpen
  \bibfield  {author} {\bibinfo {author} {\bibfnamefont {S.}~\bibnamefont
  {Arnold}}, \bibinfo {author} {\bibfnamefont {A.}~\bibnamefont {Efremov}},
  \bibinfo {author} {\bibfnamefont {K.}~\bibnamefont {Goeke}}, \bibinfo
  {author} {\bibfnamefont {M.}~\bibnamefont {Schlegel}}, \ and\ \bibinfo
  {author} {\bibfnamefont {P.}~\bibnamefont {Schweitzer}},\ }\href@noop {} {
  (\bibinfo {year} {2008})},\ \Eprint {http://arxiv.org/abs/0805.2137}
  {arXiv:0805.2137 [hep-ph]} \BibitemShut {NoStop}%
%%CITATION = ARXIV:0805.2137;%%
\bibitem [{\citenamefont {Anselmino}\ \emph
  {et~al.}(2009){\natexlab{a}}\citenamefont {Anselmino} \emph
  {et~al.}}]{Anselmino:2008sga}%
  \BibitemOpen
  \bibfield  {author} {\bibinfo {author} {\bibfnamefont {M.}~\bibnamefont
  {Anselmino}} \emph {et~al.},\ }\Doi {10.1140/epja/i2008-10697-y} {\bibfield
  {journal} {\bibinfo  {journal} {Eur. Phys. J.},\ }\textbf {\bibinfo {volume}
  {A39}},\ \bibinfo {pages} {89} (\bibinfo {year} {2009}{\natexlab{a}})},\
  \Eprint {http://arxiv.org/abs/0805.2677} {arXiv:0805.2677 [hep-ph]}
  \BibitemShut {NoStop}%
%%CITATION = 0805.2677;%%
\bibitem [{\citenamefont {Ji}\ \emph {et~al.}(2006)\citenamefont {Ji},
  \citenamefont {Qiu}, \citenamefont {Vogelsang},\ and\ \citenamefont
  {Yuan}}]{Ji:2006ub}%
  \BibitemOpen
  \bibfield  {author} {\bibinfo {author} {\bibfnamefont {X.}~\bibnamefont
  {Ji}}, \bibinfo {author} {\bibfnamefont {J.-W.}\ \bibnamefont {Qiu}},
  \bibinfo {author} {\bibfnamefont {W.}~\bibnamefont {Vogelsang}}, \ and\
  \bibinfo {author} {\bibfnamefont {F.}~\bibnamefont {Yuan}},\ }\Doi
  {10.1103/PhysRevLett.97.082002} {\bibfield  {journal} {\bibinfo  {journal}
  {Phys. Rev. Lett.},\ }\textbf {\bibinfo {volume} {97}},\ \bibinfo {pages}
  {082002} (\bibinfo {year} {2006})},\ \Eprint
  {http://arxiv.org/abs/hep-ph/0602239} {arXiv:hep-ph/0602239} \BibitemShut
  {NoStop}%
%%CITATION = HEP-PH/0602239;%%
\bibitem [{\citenamefont {Koike}\ \emph {et~al.}(2008)\citenamefont {Koike},
  \citenamefont {Vogelsang},\ and\ \citenamefont {Yuan}}]{Koike:2007dg}%
  \BibitemOpen
  \bibfield  {author} {\bibinfo {author} {\bibfnamefont {Y.}~\bibnamefont
  {Koike}}, \bibinfo {author} {\bibfnamefont {W.}~\bibnamefont {Vogelsang}}, \
  and\ \bibinfo {author} {\bibfnamefont {F.}~\bibnamefont {Yuan}},\ }\Doi
  {10.1016/j.physletb.2007.11.096} {\bibfield  {journal} {\bibinfo  {journal}
  {Phys. Lett.},\ }\textbf {\bibinfo {volume} {B659}},\ \bibinfo {pages} {878}
  (\bibinfo {year} {2008})},\ \Eprint {http://arxiv.org/abs/0711.0636}
  {arXiv:0711.0636 [hep-ph]} \BibitemShut {NoStop}%
%%CITATION = 0711.0636;%%
\bibitem [{\citenamefont {Bacchetta}\ \emph {et~al.}(2008)\citenamefont
  {Bacchetta}, \citenamefont {Boer}, \citenamefont {Diehl},\ and\ \citenamefont
  {Mulders}}]{Bacchetta:2008xw}%
  \BibitemOpen
  \bibfield  {author} {\bibinfo {author} {\bibfnamefont {A.}~\bibnamefont
  {Bacchetta}}, \bibinfo {author} {\bibfnamefont {D.}~\bibnamefont {Boer}},
  \bibinfo {author} {\bibfnamefont {M.}~\bibnamefont {Diehl}}, \ and\ \bibinfo
  {author} {\bibfnamefont {P.~J.}\ \bibnamefont {Mulders}},\ }\Doi
  {10.1088/1126-6708/2008/08/023} {\bibfield  {journal} {\bibinfo  {journal}
  {JHEP},\ }\textbf {\bibinfo {volume} {08}},\ \bibinfo {pages} {023} (\bibinfo
  {year} {2008})},\ \Eprint {http://arxiv.org/abs/0803.0227} {arXiv:0803.0227
  [hep-ph]} \BibitemShut {NoStop}%
%%CITATION = 0803.0227;%%
\bibitem [{\citenamefont {Boer}\ \emph {et~al.}(2003)\citenamefont {Boer},
  \citenamefont {Mulders},\ and\ \citenamefont {Pijlman}}]{Boer:2003cm}%
  \BibitemOpen
  \bibfield  {author} {\bibinfo {author} {\bibfnamefont {D.}~\bibnamefont
  {Boer}}, \bibinfo {author} {\bibfnamefont {P.~J.}\ \bibnamefont {Mulders}}, \
  and\ \bibinfo {author} {\bibfnamefont {F.}~\bibnamefont {Pijlman}},\ }\Doi
  {10.1016/S0550-3213(03)00527-3} {\bibfield  {journal} {\bibinfo  {journal}
  {Nucl. Phys.},\ }\textbf {\bibinfo {volume} {B667}},\ \bibinfo {pages} {201}
  (\bibinfo {year} {2003})},\ \Eprint {http://arxiv.org/abs/hep-ph/0303034}
  {arXiv:hep-ph/0303034} \BibitemShut {NoStop}%
%%CITATION = HEP-PH/0303034;%%
\bibitem [{\citenamefont {Ma}\ and\ \citenamefont {Wang}(2004)}]{Ma:2003ut}%
  \BibitemOpen
  \bibfield  {author} {\bibinfo {author} {\bibfnamefont {J.~P.}\ \bibnamefont
  {Ma}}\ and\ \bibinfo {author} {\bibfnamefont {Q.}~\bibnamefont {Wang}},\
  }\Doi {10.1140/epjc/s2004-02009-x} {\bibfield  {journal} {\bibinfo  {journal}
  {Eur. Phys. J.},\ }\textbf {\bibinfo {volume} {C37}},\ \bibinfo {pages} {293}
  (\bibinfo {year} {2004})},\ \Eprint {http://arxiv.org/abs/hep-ph/0310245}
  {arXiv:hep-ph/0310245} \BibitemShut {NoStop}%
%%CITATION = HEP-PH/0310245;%%
\bibitem [{\citenamefont {Kang}\ \emph
  {et~al.}(2011){\natexlab{a}}\citenamefont {Kang}, \citenamefont {Qiu},
  \citenamefont {Vogelsang},\ and\ \citenamefont {Yuan}}]{Kang:2011hk}%
  \BibitemOpen
  \bibfield  {author} {\bibinfo {author} {\bibfnamefont {Z.-B.}\ \bibnamefont
  {Kang}}, \bibinfo {author} {\bibfnamefont {J.-W.}\ \bibnamefont {Qiu}},
  \bibinfo {author} {\bibfnamefont {W.}~\bibnamefont {Vogelsang}}, \ and\
  \bibinfo {author} {\bibfnamefont {F.}~\bibnamefont {Yuan}},\ }\Doi
  {10.1103/PhysRevD.83.094001} {\bibfield  {journal} {\bibinfo  {journal}
  {Phys. Rev.},\ }\textbf {\bibinfo {volume} {D83}},\ \bibinfo {pages} {094001}
  (\bibinfo {year} {2011}{\natexlab{a}})},\ \Eprint
  {http://arxiv.org/abs/1103.1591} {arXiv:1103.1591 [hep-ph]} \BibitemShut
  {NoStop}%
%%CITATION = 1103.1591;%%
\bibitem [{\citenamefont {Collins}(2011)}]{collins}%
  \BibitemOpen
  \bibfield  {author} {\bibinfo {author} {\bibfnamefont {J.~C.}\ \bibnamefont
  {Collins}},\ }\href@noop {} {\emph {\bibinfo {title} {Foundations of
  Perturbative QCD}}}\ (\bibinfo  {publisher} {Cambridge University Press},\
  \bibinfo {address} {Cambridge},\ \bibinfo {year} {2011})\BibitemShut
  {NoStop}%
\bibitem [{\citenamefont {Collins}(2002)}]{Collins:2002kn}%
  \BibitemOpen
  \bibfield  {author} {\bibinfo {author} {\bibfnamefont {J.~C.}\ \bibnamefont
  {Collins}},\ }\href@noop {} {\bibfield  {journal} {\bibinfo  {journal} {Phys.
  Lett.},\ }\textbf {\bibinfo {volume} {B536}},\ \bibinfo {pages} {43}
  (\bibinfo {year} {2002})},\ \Eprint {http://arxiv.org/abs/hep-ph/0204004}
  {hep-ph/0204004} \BibitemShut {NoStop}%
%%CITATION = HEP-PH/0204004;%%
\bibitem [{\citenamefont {Collins}\ \emph {et~al.}(2006)\citenamefont {Collins}
  \emph {et~al.}}]{Collins:2005rq}%
  \BibitemOpen
  \bibfield  {author} {\bibinfo {author} {\bibfnamefont {J.~C.}\ \bibnamefont
  {Collins}} \emph {et~al.},\ }\Doi {10.1103/PhysRevD.73.094023} {\bibfield
  {journal} {\bibinfo  {journal} {Phys. Rev.},\ }\textbf {\bibinfo {volume}
  {D73}},\ \bibinfo {pages} {094023} (\bibinfo {year} {2006})},\ \Eprint
  {http://arxiv.org/abs/hep-ph/0511272} {arXiv:hep-ph/0511272} \BibitemShut
  {NoStop}%
%%CITATION = HEP-PH/0511272;%%
\bibitem [{\citenamefont {Kang}\ and\ \citenamefont {Qiu}(2010)}]{Kang:2009sm}%
  \BibitemOpen
  \bibfield  {author} {\bibinfo {author} {\bibfnamefont {Z.-B.}\ \bibnamefont
  {Kang}}\ and\ \bibinfo {author} {\bibfnamefont {J.-W.}\ \bibnamefont {Qiu}},\
  }\Doi {10.1103/PhysRevD.81.054020} {\bibfield  {journal} {\bibinfo  {journal}
  {Phys.Rev.},\ }\textbf {\bibinfo {volume} {D81}},\ \bibinfo {pages} {054020}
  (\bibinfo {year} {2010})},\ \Eprint {http://arxiv.org/abs/0912.1319}
  {arXiv:0912.1319 [hep-ph]} \BibitemShut {NoStop}%
%%CITATION = ARXIV:0912.1319;%%
\bibitem [{\citenamefont {Anselmino}\ \emph
  {et~al.}(2009){\natexlab{b}}\citenamefont {Anselmino} \emph
  {et~al.}}]{Anselmino:2009st}%
  \BibitemOpen
  \bibfield  {author} {\bibinfo {author} {\bibfnamefont {M.}~\bibnamefont
  {Anselmino}} \emph {et~al.},\ }\Doi {10.1103/PhysRevD.79.054010} {\bibfield
  {journal} {\bibinfo  {journal} {Phys. Rev.},\ }\textbf {\bibinfo {volume}
  {D79}},\ \bibinfo {pages} {054010} (\bibinfo {year} {2009}{\natexlab{b}})},\
  \Eprint {http://arxiv.org/abs/0901.3078} {arXiv:0901.3078 [hep-ph]}
  \BibitemShut {NoStop}%
%%CITATION = 0901.3078;%%
\bibitem [{\citenamefont {Kang}\ \emph
  {et~al.}(2011){\natexlab{b}}\citenamefont {Kang}, \citenamefont {Xiao},\ and\
  \citenamefont {Yuan}}]{Kang:2011mr}%
  \BibitemOpen
  \bibfield  {author} {\bibinfo {author} {\bibfnamefont {Z.-B.}\ \bibnamefont
  {Kang}}, \bibinfo {author} {\bibfnamefont {B.-W.}\ \bibnamefont {Xiao}}, \
  and\ \bibinfo {author} {\bibfnamefont {F.}~\bibnamefont {Yuan}},\ }\Doi
  {10.1103/PhysRevLett.107.152002} {\bibfield  {journal} {\bibinfo  {journal}
  {Phys. Rev. Lett.},\ }\textbf {\bibinfo {volume} {107}},\ \bibinfo {pages}
  {152002} (\bibinfo {year} {2011}{\natexlab{b}})},\ \Eprint
  {http://arxiv.org/abs/1106.0266} {arXiv:1106.0266 [hep-ph]} \BibitemShut
  {NoStop}%
%%CITATION = 1106.0266;%%
\bibitem [{\citenamefont {Aybat}\ \emph
  {et~al.}(2011){\natexlab{a}}\citenamefont {Aybat}, \citenamefont {Collins},
  \citenamefont {Qiu},\ and\ \citenamefont {Rogers}}]{Aybat:2011ge}%
  \BibitemOpen
  \bibfield  {author} {\bibinfo {author} {\bibfnamefont {S.~M.}\ \bibnamefont
  {Aybat}}, \bibinfo {author} {\bibfnamefont {J.~C.}\ \bibnamefont {Collins}},
  \bibinfo {author} {\bibfnamefont {J.-W.}\ \bibnamefont {Qiu}}, \ and\
  \bibinfo {author} {\bibfnamefont {T.~C.}\ \bibnamefont {Rogers}},\
  }\href@noop {} { (\bibinfo {year} {2011}{\natexlab{a}})},\ \Eprint
  {http://arxiv.org/abs/1110.6428} {arXiv:1110.6428 [hep-ph]} \BibitemShut
  {NoStop}%
%%CITATION = 1110.6428;%%
\bibitem [{\citenamefont {Metz}\ and\ \citenamefont
  {Zhou}(2011)}]{Metz:2010xs}%
  \BibitemOpen
  \bibfield  {author} {\bibinfo {author} {\bibfnamefont {A.}~\bibnamefont
  {Metz}}\ and\ \bibinfo {author} {\bibfnamefont {J.}~\bibnamefont {Zhou}},\
  }\Doi {10.1016/j.physletb.2011.04.042} {\bibfield  {journal} {\bibinfo
  {journal} {Phys.Lett.},\ }\textbf {\bibinfo {volume} {B700}},\ \bibinfo
  {pages} {11} (\bibinfo {year} {2011})},\ \Eprint
  {http://arxiv.org/abs/1006.3097} {arXiv:1006.3097 [hep-ph]} \BibitemShut
  {NoStop}%
%%CITATION = ARXIV:1006.3097;%%
\bibitem [{\citenamefont {Kang}\ and\ \citenamefont
  {Qiu}(2009){\natexlab{a}}}]{Kang:2009bp}%
  \BibitemOpen
  \bibfield  {author} {\bibinfo {author} {\bibfnamefont {Z.-B.}\ \bibnamefont
  {Kang}}\ and\ \bibinfo {author} {\bibfnamefont {J.-W.}\ \bibnamefont {Qiu}},\
  }\Doi {10.1103/PhysRevLett.103.172001} {\bibfield  {journal} {\bibinfo
  {journal} {Phys.Rev.Lett.},\ }\textbf {\bibinfo {volume} {103}},\ \bibinfo
  {pages} {172001} (\bibinfo {year} {2009}{\natexlab{a}})},\ \Eprint
  {http://arxiv.org/abs/0903.3629} {arXiv:0903.3629 [hep-ph]} \BibitemShut
  {NoStop}%
%%CITATION = ARXIV:0903.3629;%%
\bibitem [{\citenamefont {Anselmino}\ \emph {et~al.}(2010)\citenamefont
  {Anselmino} \emph {et~al.}}]{Anselmino:2009pn}%
  \BibitemOpen
  \bibfield  {author} {\bibinfo {author} {\bibfnamefont {M.}~\bibnamefont
  {Anselmino}} \emph {et~al.},\ }\Doi {10.1103/PhysRevD.81.034007} {\bibfield
  {journal} {\bibinfo  {journal} {Phys. Rev.},\ }\textbf {\bibinfo {volume}
  {D81}},\ \bibinfo {pages} {034007} (\bibinfo {year} {2010})},\ \Eprint
  {http://arxiv.org/abs/0911.1744} {arXiv:0911.1744 [hep-ph]} \BibitemShut
  {NoStop}%
%%CITATION = 0911.1744;%%
\bibitem [{\citenamefont {Kang}\ \emph
  {et~al.}(2011){\natexlab{c}}\citenamefont {Kang}, \citenamefont {Metz},
  \citenamefont {Qiu},\ and\ \citenamefont {Zhou}}]{Kang:2011jw}%
  \BibitemOpen
  \bibfield  {author} {\bibinfo {author} {\bibfnamefont {Z.-B.}\ \bibnamefont
  {Kang}}, \bibinfo {author} {\bibfnamefont {A.}~\bibnamefont {Metz}}, \bibinfo
  {author} {\bibfnamefont {J.-W.}\ \bibnamefont {Qiu}}, \ and\ \bibinfo
  {author} {\bibfnamefont {J.}~\bibnamefont {Zhou}},\ }\Doi
  {10.1103/PhysRevD.84.034046} {\bibfield  {journal} {\bibinfo  {journal}
  {Phys. Rev.},\ }\textbf {\bibinfo {volume} {D84}},\ \bibinfo {pages} {034046}
  (\bibinfo {year} {2011}{\natexlab{c}})},\ \Eprint
  {http://arxiv.org/abs/1106.3514} {arXiv:1106.3514 [hep-ph]} \BibitemShut
  {NoStop}%
%%CITATION = 1106.3514;%%
\bibitem [{\citenamefont {Contalbrigo}\ \emph {et~al.}(2010)\citenamefont
  {Contalbrigo}, \citenamefont {Lopez~Ruiz}, \citenamefont {Manfre},
  \citenamefont {Martinez de~la Ossa},\ and\ \citenamefont
  {Sanftl}}]{Contalbrigo:2010zz}%
  \BibitemOpen
  \bibfield  {author} {\bibinfo {author} {\bibfnamefont {M.}~\bibnamefont
  {Contalbrigo}}, \bibinfo {author} {\bibfnamefont {A.}~\bibnamefont
  {Lopez~Ruiz}}, \bibinfo {author} {\bibfnamefont {L.}~\bibnamefont {Manfre}},
  \bibinfo {author} {\bibfnamefont {A.}~\bibnamefont {Martinez de~la Ossa}}, \
  and\ \bibinfo {author} {\bibfnamefont {F.}~\bibnamefont {Sanftl}} (\bibinfo
  {collaboration} {HERMES}),\ }\href@noop {} {\bibfield  {journal} {\bibinfo
  {journal} {PoS},\ }\textbf {\bibinfo {volume} {DIS2010}},\ \bibinfo {pages}
  {236} (\bibinfo {year} {2010})}\BibitemShut {NoStop}%
%%CITATION = POSCI,DIS2010,236;%%
\bibitem [{\citenamefont {Yuan}(2008)}]{Yuan:2007nd}%
  \BibitemOpen
  \bibfield  {author} {\bibinfo {author} {\bibfnamefont {F.}~\bibnamefont
  {Yuan}},\ }\Doi {10.1103/PhysRevLett.100.032003} {\bibfield  {journal}
  {\bibinfo  {journal} {Phys. Rev. Lett.},\ }\textbf {\bibinfo {volume}
  {100}},\ \bibinfo {pages} {032003} (\bibinfo {year} {2008})},\ \Eprint
  {http://arxiv.org/abs/0709.3272} {arXiv:0709.3272 [hep-ph]} \BibitemShut
  {NoStop}%
%%CITATION = 0709.3272;%%
\bibitem [{\citenamefont {D'Alesio}\ \emph
  {et~al.}(2011){\natexlab{a}}\citenamefont {D'Alesio}, \citenamefont
  {Murgia},\ and\ \citenamefont {Pisano}}]{D'Alesio:2010am}%
  \BibitemOpen
  \bibfield  {author} {\bibinfo {author} {\bibfnamefont {U.}~\bibnamefont
  {D'Alesio}}, \bibinfo {author} {\bibfnamefont {F.}~\bibnamefont {Murgia}}, \
  and\ \bibinfo {author} {\bibfnamefont {C.}~\bibnamefont {Pisano}},\ }\Doi
  {10.1103/PhysRevD.83.034021} {\bibfield  {journal} {\bibinfo  {journal}
  {Phys.Rev.},\ }\textbf {\bibinfo {volume} {D83}},\ \bibinfo {pages} {034021}
  (\bibinfo {year} {2011}{\natexlab{a}})},\ \Eprint
  {http://arxiv.org/abs/1011.2692} {arXiv:1011.2692 [hep-ph]} \BibitemShut
  {NoStop}%
%%CITATION = ARXIV:1011.2692;%%
\bibitem [{\citenamefont {D'Alesio}\ \emph
  {et~al.}(2011){\natexlab{b}}\citenamefont {D'Alesio}, \citenamefont
  {Gamberg}, \citenamefont {Kang}, \citenamefont {Murgia},\ and\ \citenamefont
  {Pisano}}]{D'Alesio:2011mc}%
  \BibitemOpen
  \bibfield  {author} {\bibinfo {author} {\bibfnamefont {U.}~\bibnamefont
  {D'Alesio}}, \bibinfo {author} {\bibfnamefont {L.}~\bibnamefont {Gamberg}},
  \bibinfo {author} {\bibfnamefont {Z.-B.}\ \bibnamefont {Kang}}, \bibinfo
  {author} {\bibfnamefont {F.}~\bibnamefont {Murgia}}, \ and\ \bibinfo {author}
  {\bibfnamefont {C.}~\bibnamefont {Pisano}},\ }\Doi
  {10.1016/j.physletb.2011.09.067} {\bibfield  {journal} {\bibinfo  {journal}
  {Phys.Lett.},\ }\textbf {\bibinfo {volume} {B704}},\ \bibinfo {pages} {637}
  (\bibinfo {year} {2011}{\natexlab{b}})},\ \Eprint
  {http://arxiv.org/abs/1108.0827} {arXiv:1108.0827 [hep-ph]} \BibitemShut
  {NoStop}%
%%CITATION = ARXIV:1108.0827;%%
\bibitem [{\citenamefont {Fersch}(2011)}]{Fersch:2011zz}%
  \BibitemOpen
  \bibfield  {author} {\bibinfo {author} {\bibfnamefont {R.}~\bibnamefont
  {Fersch}} (\bibinfo {collaboration} {STAR Collaboration}),\ }\Doi
  {10.1088/1742-6596/295/1/012048} {\bibfield  {journal} {\bibinfo  {journal}
  {J.Phys.Conf.Ser.},\ }\textbf {\bibinfo {volume} {295}},\ \bibinfo {pages}
  {012048} (\bibinfo {year} {2011})}\BibitemShut {NoStop}%
%%CITATION = 00462,295,012048;%%
\bibitem [{\citenamefont {Boer}(2011)}]{Boer:2011fx}%
  \BibitemOpen
  \bibfield  {author} {\bibinfo {author} {\bibfnamefont {D.}~\bibnamefont
  {Boer}},\ }\Doi {10.1016/j.physletb.2011.07.006} {\bibfield  {journal}
  {\bibinfo  {journal} {Phys. Lett.},\ }\textbf {\bibinfo {volume} {B702}},\
  \bibinfo {pages} {242} (\bibinfo {year} {2011})},\ \Eprint
  {http://arxiv.org/abs/1105.2543} {arXiv:1105.2543 [hep-ph]} \BibitemShut
  {NoStop}%
%%CITATION = 1105.2543;%%
\bibitem [{\citenamefont {Bacchetta}\ \emph {et~al.}(2007)\citenamefont
  {Bacchetta} \emph {et~al.}}]{Bacchetta:2006tn}%
  \BibitemOpen
  \bibfield  {author} {\bibinfo {author} {\bibfnamefont {A.}~\bibnamefont
  {Bacchetta}} \emph {et~al.},\ }\href@noop {} {\bibfield  {journal} {\bibinfo
  {journal} {JHEP},\ }\textbf {\bibinfo {volume} {02}},\ \bibinfo {pages} {093}
  (\bibinfo {year} {2007})},\ \Eprint {http://arxiv.org/abs/hep-ph/0611265}
  {hep-ph/0611265} \BibitemShut {NoStop}%
%%CITATION = HEP-PH/0611265;%%
\bibitem [{\citenamefont {Mulders}\ and\ \citenamefont
  {Tangerman}(1996)}]{Mulders:1995dh}%
  \BibitemOpen
  \bibfield  {author} {\bibinfo {author} {\bibfnamefont {P.~J.}\ \bibnamefont
  {Mulders}}\ and\ \bibinfo {author} {\bibfnamefont {R.~D.}\ \bibnamefont
  {Tangerman}},\ }\Doi {10.1016/0550-3213(95)00632-X} {\bibfield  {journal}
  {\bibinfo  {journal} {Nucl. Phys.},\ }\textbf {\bibinfo {volume} {B461}},\
  \bibinfo {pages} {197} (\bibinfo {year} {1996})},\ \Eprint
  {http://arxiv.org/abs/hep-ph/9510301} {arXiv:hep-ph/9510301} \BibitemShut
  {NoStop}%
%%CITATION = HEP-PH/9510301;%%
\bibitem [{\citenamefont {Bacchetta}\ \emph {et~al.}(2004)\citenamefont
  {Bacchetta}, \citenamefont {D'Alesio}, \citenamefont {Diehl},\ and\
  \citenamefont {Miller}}]{Bacchetta:2004jz}%
  \BibitemOpen
  \bibfield  {author} {\bibinfo {author} {\bibfnamefont {A.}~\bibnamefont
  {Bacchetta}}, \bibinfo {author} {\bibfnamefont {U.}~\bibnamefont {D'Alesio}},
  \bibinfo {author} {\bibfnamefont {M.}~\bibnamefont {Diehl}}, \ and\ \bibinfo
  {author} {\bibfnamefont {C.~A.}\ \bibnamefont {Miller}},\ }\href@noop {}
  {\bibfield  {journal} {\bibinfo  {journal} {Phys. Rev.},\ }\textbf {\bibinfo
  {volume} {D70}},\ \bibinfo {pages} {117504} (\bibinfo {year} {2004})},\
  \Eprint {http://arxiv.org/abs/hep-ph/0410050} {hep-ph/0410050} \BibitemShut
  {NoStop}%
%%CITATION = HEP-PH 0410050;%%
\bibitem [{\citenamefont {Airapetian}\ \emph {et~al.}(2009)\citenamefont
  {Airapetian} \emph {et~al.}}]{:2009ti}%
  \BibitemOpen
  \bibfield  {author} {\bibinfo {author} {\bibfnamefont {A.}~\bibnamefont
  {Airapetian}} \emph {et~al.} (\bibinfo {collaboration} {HERMES}),\ }\Doi
  {10.1103/PhysRevLett.103.152002} {\bibfield  {journal} {\bibinfo  {journal}
  {Phys. Rev. Lett.},\ }\textbf {\bibinfo {volume} {103}},\ \bibinfo {pages}
  {152002} (\bibinfo {year} {2009})},\ \Eprint {http://arxiv.org/abs/0906.3918}
  {arXiv:0906.3918 [hep-ex]} \BibitemShut {NoStop}%
%%CITATION = 0906.3918;%%
\bibitem [{\citenamefont {Alekseev}\ \emph {et~al.}(2009)\citenamefont
  {Alekseev} \emph {et~al.}}]{:2008dn}%
  \BibitemOpen
  \bibfield  {author} {\bibinfo {author} {\bibfnamefont {M.}~\bibnamefont
  {Alekseev}} \emph {et~al.} (\bibinfo {collaboration} {COMPASS}),\ }\Doi
  {10.1016/j.physletb.2009.01.060} {\bibfield  {journal} {\bibinfo  {journal}
  {Phys. Lett.},\ }\textbf {\bibinfo {volume} {B673}},\ \bibinfo {pages} {127}
  (\bibinfo {year} {2009})},\ \Eprint {http://arxiv.org/abs/0802.2160}
  {arXiv:0802.2160 [hep-ex]} \BibitemShut {NoStop}%
%%CITATION = 0802.2160;%%
\bibitem [{\citenamefont {Owens}(1987)}]{Owens:1986mp}%
  \BibitemOpen
  \bibfield  {author} {\bibinfo {author} {\bibfnamefont {J.~F.}\ \bibnamefont
  {Owens}},\ }\Doi {10.1103/RevModPhys.59.465} {\bibfield  {journal} {\bibinfo
  {journal} {Rev. Mod. Phys.},\ }\textbf {\bibinfo {volume} {59}},\ \bibinfo
  {pages} {465} (\bibinfo {year} {1987})}\BibitemShut {NoStop}%
%%CITATION = RMPHA,59,465;%%
\bibitem [{\citenamefont {Kang}\ and\ \citenamefont
  {Yuan}(2010)}]{Kang:2010vd}%
  \BibitemOpen
  \bibfield  {author} {\bibinfo {author} {\bibfnamefont {Z.-B.}\ \bibnamefont
  {Kang}}\ and\ \bibinfo {author} {\bibfnamefont {F.}~\bibnamefont {Yuan}},\
  }\Doi {10.1103/PhysRevD.81.054007} {\bibfield  {journal} {\bibinfo  {journal}
  {Phys.Rev.},\ }\textbf {\bibinfo {volume} {D81}},\ \bibinfo {pages} {054007}
  (\bibinfo {year} {2010})},\ \Eprint {http://arxiv.org/abs/1001.0247}
  {arXiv:1001.0247 [hep-ph]} \BibitemShut {NoStop}%
%%CITATION = ARXIV:1001.0247;%%
\bibitem [{\citenamefont {Gamberg}\ and\ \citenamefont
  {Kang}(2011)}]{Gamberg:2010tj}%
  \BibitemOpen
  \bibfield  {author} {\bibinfo {author} {\bibfnamefont {L.}~\bibnamefont
  {Gamberg}}\ and\ \bibinfo {author} {\bibfnamefont {Z.-B.}\ \bibnamefont
  {Kang}},\ }\Doi {10.1016/j.physletb.2010.11.066} {\bibfield  {journal}
  {\bibinfo  {journal} {Phys. Lett.},\ }\textbf {\bibinfo {volume} {B696}},\
  \bibinfo {pages} {109} (\bibinfo {year} {2011})},\ \Eprint
  {http://arxiv.org/abs/1009.1936} {arXiv:1009.1936 [hep-ph]} \BibitemShut
  {NoStop}%
%%CITATION = 1009.1936;%%
\bibitem [{\citenamefont {de~Florian}\ \emph {et~al.}(2008)\citenamefont
  {de~Florian}, \citenamefont {Sassot}, \citenamefont {Stratmann},\ and\
  \citenamefont {Vogelsang}}]{deFlorian:2008mr}%
  \BibitemOpen
  \bibfield  {author} {\bibinfo {author} {\bibfnamefont {D.}~\bibnamefont
  {de~Florian}}, \bibinfo {author} {\bibfnamefont {R.}~\bibnamefont {Sassot}},
  \bibinfo {author} {\bibfnamefont {M.}~\bibnamefont {Stratmann}}, \ and\
  \bibinfo {author} {\bibfnamefont {W.}~\bibnamefont {Vogelsang}},\ }\Doi
  {10.1103/PhysRevLett.101.072001} {\bibfield  {journal} {\bibinfo  {journal}
  {Phys. Rev. Lett.},\ }\textbf {\bibinfo {volume} {101}},\ \bibinfo {pages}
  {072001} (\bibinfo {year} {2008})},\ \Eprint {http://arxiv.org/abs/0804.0422}
  {arXiv:0804.0422 [hep-ph]} \BibitemShut {NoStop}%
%%CITATION = 0804.0422;%%
\bibitem [{\citenamefont {de~Florian}\ \emph {et~al.}(2009)\citenamefont
  {de~Florian}, \citenamefont {Sassot}, \citenamefont {Stratmann},\ and\
  \citenamefont {Vogelsang}}]{deFlorian:2009vb}%
  \BibitemOpen
  \bibfield  {author} {\bibinfo {author} {\bibfnamefont {D.}~\bibnamefont
  {de~Florian}}, \bibinfo {author} {\bibfnamefont {R.}~\bibnamefont {Sassot}},
  \bibinfo {author} {\bibfnamefont {M.}~\bibnamefont {Stratmann}}, \ and\
  \bibinfo {author} {\bibfnamefont {W.}~\bibnamefont {Vogelsang}},\ }\Doi
  {10.1103/PhysRevD.80.034030} {\bibfield  {journal} {\bibinfo  {journal}
  {Phys.Rev.},\ }\textbf {\bibinfo {volume} {D80}},\ \bibinfo {pages} {034030}
  (\bibinfo {year} {2009})},\ \Eprint {http://arxiv.org/abs/0904.3821}
  {arXiv:0904.3821 [hep-ph]} \BibitemShut {NoStop}%
%%CITATION = ARXIV:0904.3821;%%
\bibitem [{\citenamefont {Abelev}\ \emph {et~al.}(2008)\citenamefont {Abelev}
  \emph {et~al.}}]{:2008qb}%
  \BibitemOpen
  \bibfield  {author} {\bibinfo {author} {\bibfnamefont {B.~I.}\ \bibnamefont
  {Abelev}} \emph {et~al.} (\bibinfo {collaboration} {STAR}),\ }\Doi
  {10.1103/PhysRevLett.101.222001} {\bibfield  {journal} {\bibinfo  {journal}
  {Phys. Rev. Lett.},\ }\textbf {\bibinfo {volume} {101}},\ \bibinfo {pages}
  {222001} (\bibinfo {year} {2008})},\ \Eprint {http://arxiv.org/abs/0801.2990}
  {arXiv:0801.2990 [hep-ex]} \BibitemShut {NoStop}%
%%CITATION = 0801.2990;%%
\bibitem [{\citenamefont {Lee}\ and\ \citenamefont
  {Videbaek}(2007)}]{Lee:2007zzh}%
  \BibitemOpen
  \bibfield  {author} {\bibinfo {author} {\bibfnamefont {J.~H.}\ \bibnamefont
  {Lee}}\ and\ \bibinfo {author} {\bibfnamefont {F.}~\bibnamefont {Videbaek}}
  (\bibinfo {collaboration} {BRAHMS}),\ }\href@noop {} {\bibfield  {journal}
  {\bibinfo  {journal} {AIP Conf. Proc.},\ }\textbf {\bibinfo {volume} {915}},\
  \bibinfo {pages} {533} (\bibinfo {year} {2007})}\BibitemShut {NoStop}%
%%CITATION = APCPC,915,533;%%
\bibitem [{\citenamefont {Gluck}\ \emph {et~al.}(1998)\citenamefont {Gluck},
  \citenamefont {Reya},\ and\ \citenamefont {Vogt}}]{Gluck:1998xa}%
  \BibitemOpen
  \bibfield  {author} {\bibinfo {author} {\bibfnamefont {M.}~\bibnamefont
  {Gluck}}, \bibinfo {author} {\bibfnamefont {E.}~\bibnamefont {Reya}}, \ and\
  \bibinfo {author} {\bibfnamefont {A.}~\bibnamefont {Vogt}},\ }\Doi
  {10.1007/s100520050289} {\bibfield  {journal} {\bibinfo  {journal} {Eur.
  Phys. J.},\ }\textbf {\bibinfo {volume} {C5}},\ \bibinfo {pages} {461}
  (\bibinfo {year} {1998})},\ \Eprint {http://arxiv.org/abs/hep-ph/9806404}
  {arXiv:hep-ph/9806404} \BibitemShut {NoStop}%
%%CITATION = HEP-PH/9806404;%%
\bibitem [{\citenamefont {de~Florian}\ \emph {et~al.}(2007)\citenamefont
  {de~Florian}, \citenamefont {Sassot},\ and\ \citenamefont
  {Stratmann}}]{deFlorian:2007aj}%
  \BibitemOpen
  \bibfield  {author} {\bibinfo {author} {\bibfnamefont {D.}~\bibnamefont
  {de~Florian}}, \bibinfo {author} {\bibfnamefont {R.}~\bibnamefont {Sassot}},
  \ and\ \bibinfo {author} {\bibfnamefont {M.}~\bibnamefont {Stratmann}},\
  }\href@noop {} {\bibfield  {journal} {\bibinfo  {journal} {Phys. Rev.},\
  }\textbf {\bibinfo {volume} {D75}},\ \bibinfo {pages} {114010} (\bibinfo
  {year} {2007})},\ \Eprint {http://arxiv.org/abs/hep-ph/0703242}
  {hep-ph/0703242} \BibitemShut {NoStop}%
%%CITATION = HEP-PH/0703242;%%
\bibitem [{\citenamefont {Vogelsang}\ and\ \citenamefont
  {Yuan}(2009)}]{Vogelsang:2009pj}%
  \BibitemOpen
  \bibfield  {author} {\bibinfo {author} {\bibfnamefont {W.}~\bibnamefont
  {Vogelsang}}\ and\ \bibinfo {author} {\bibfnamefont {F.}~\bibnamefont
  {Yuan}},\ }\Doi {10.1103/PhysRevD.79.094010} {\bibfield  {journal} {\bibinfo
  {journal} {Phys. Rev.},\ }\textbf {\bibinfo {volume} {D79}},\ \bibinfo
  {pages} {094010} (\bibinfo {year} {2009})},\ \Eprint
  {http://arxiv.org/abs/0904.0410} {arXiv:0904.0410 [hep-ph]} \BibitemShut
  {NoStop}%
%%CITATION = 0904.0410;%%
\bibitem [{\citenamefont {Kang}\ and\ \citenamefont
  {Qiu}(2009){\natexlab{b}}}]{Kang:2008ey}%
  \BibitemOpen
  \bibfield  {author} {\bibinfo {author} {\bibfnamefont {Z.-B.}\ \bibnamefont
  {Kang}}\ and\ \bibinfo {author} {\bibfnamefont {J.-W.}\ \bibnamefont {Qiu}},\
  }\Doi {10.1103/PhysRevD.79.016003} {\bibfield  {journal} {\bibinfo  {journal}
  {Phys.Rev.},\ }\textbf {\bibinfo {volume} {D79}},\ \bibinfo {pages} {016003}
  (\bibinfo {year} {2009}{\natexlab{b}})},\ \Eprint
  {http://arxiv.org/abs/0811.3101} {arXiv:0811.3101 [hep-ph]} \BibitemShut
  {NoStop}%
%%CITATION = ARXIV:0811.3101;%%
\bibitem [{\citenamefont {Zhou}\ \emph {et~al.}(2009)\citenamefont {Zhou},
  \citenamefont {Yuan},\ and\ \citenamefont {Liang}}]{Zhou:2008mz}%
  \BibitemOpen
  \bibfield  {author} {\bibinfo {author} {\bibfnamefont {J.}~\bibnamefont
  {Zhou}}, \bibinfo {author} {\bibfnamefont {F.}~\bibnamefont {Yuan}}, \ and\
  \bibinfo {author} {\bibfnamefont {Z.-T.}\ \bibnamefont {Liang}},\ }\Doi
  {10.1103/PhysRevD.79.114022} {\bibfield  {journal} {\bibinfo  {journal}
  {Phys.Rev.},\ }\textbf {\bibinfo {volume} {D79}},\ \bibinfo {pages} {114022}
  (\bibinfo {year} {2009})},\ \Eprint {http://arxiv.org/abs/0812.4484}
  {arXiv:0812.4484 [hep-ph]} \BibitemShut {NoStop}%
%%CITATION = ARXIV:0812.4484;%%
\bibitem [{\citenamefont {Braun}\ \emph {et~al.}(2009)\citenamefont {Braun},
  \citenamefont {Manashov},\ and\ \citenamefont {Pirnay}}]{Braun:2009mi}%
  \BibitemOpen
  \bibfield  {author} {\bibinfo {author} {\bibfnamefont {V.}~\bibnamefont
  {Braun}}, \bibinfo {author} {\bibfnamefont {A.}~\bibnamefont {Manashov}}, \
  and\ \bibinfo {author} {\bibfnamefont {B.}~\bibnamefont {Pirnay}},\ }\Doi
  {10.1103/PhysRevD.80.114002} {\bibfield  {journal} {\bibinfo  {journal}
  {Phys.Rev.},\ }\textbf {\bibinfo {volume} {D80}},\ \bibinfo {pages} {114002}
  (\bibinfo {year} {2009})},\ \Eprint {http://arxiv.org/abs/0909.3410}
  {arXiv:0909.3410 [hep-ph]} \BibitemShut {NoStop}%
%%CITATION = ARXIV:0909.3410;%%
\bibitem [{\citenamefont {Kang}(2011)}]{Kang:2010xv}%
  \BibitemOpen
  \bibfield  {author} {\bibinfo {author} {\bibfnamefont {Z.-B.}\ \bibnamefont
  {Kang}},\ }\Doi {10.1103/PhysRevD.83.036006} {\bibfield  {journal} {\bibinfo
  {journal} {Phys.Rev.},\ }\textbf {\bibinfo {volume} {D83}},\ \bibinfo {pages}
  {036006} (\bibinfo {year} {2011})},\ \Eprint {http://arxiv.org/abs/1012.3419}
  {arXiv:1012.3419 [hep-ph]} \BibitemShut {NoStop}%
%%CITATION = ARXIV:1012.3419;%%
\bibitem [{\citenamefont {Adams}\ \emph {et~al.}(1991)\citenamefont {Adams}
  \emph {et~al.}}]{Adams:1991cs}%
  \BibitemOpen
  \bibfield  {author} {\bibinfo {author} {\bibfnamefont {D.~L.}\ \bibnamefont
  {Adams}} \emph {et~al.} (\bibinfo {collaboration} {FNAL-E704}),\ }\Doi
  {10.1016/0370-2693(91)90378-4} {\bibfield  {journal} {\bibinfo  {journal}
  {Phys. Lett.},\ }\textbf {\bibinfo {volume} {B264}},\ \bibinfo {pages} {462}
  (\bibinfo {year} {1991})}\BibitemShut {NoStop}%
%%CITATION = PHLTA,B264,462;%%
\bibitem [{\citenamefont {Krueger}\ \emph {et~al.}(1999)\citenamefont
  {Krueger}, \citenamefont {Allgower}, \citenamefont {Kasprzyk}, \citenamefont
  {Spinka}, \citenamefont {Underwood} \emph {et~al.}}]{Krueger:1998hz}%
  \BibitemOpen
  \bibfield  {author} {\bibinfo {author} {\bibfnamefont {K.}~\bibnamefont
  {Krueger}}, \bibinfo {author} {\bibfnamefont {C.}~\bibnamefont {Allgower}},
  \bibinfo {author} {\bibfnamefont {T.}~\bibnamefont {Kasprzyk}}, \bibinfo
  {author} {\bibfnamefont {H.}~\bibnamefont {Spinka}}, \bibinfo {author}
  {\bibfnamefont {D.}~\bibnamefont {Underwood}},  \emph {et~al.},\ }\Doi
  {10.1016/S0370-2693(99)00677-2} {\bibfield  {journal} {\bibinfo  {journal}
  {Phys.Lett.},\ }\textbf {\bibinfo {volume} {B459}},\ \bibinfo {pages} {412}
  (\bibinfo {year} {1999})}\BibitemShut {NoStop}%
%%CITATION = PHLTA,B459,412;%%
\bibitem [{\citenamefont {Aybat}\ \emph
  {et~al.}(2011){\natexlab{b}}\citenamefont {Aybat}, \citenamefont {Prokudin},\
  and\ \citenamefont {Rogers}}]{Aybat:2011ta}%
  \BibitemOpen
  \bibfield  {author} {\bibinfo {author} {\bibfnamefont {S.~M.}\ \bibnamefont
  {Aybat}}, \bibinfo {author} {\bibfnamefont {A.}~\bibnamefont {Prokudin}}, \
  and\ \bibinfo {author} {\bibfnamefont {T.~C.}\ \bibnamefont {Rogers}},\
  }\href@noop {} { (\bibinfo {year} {2011}{\natexlab{b}})},\ \Eprint
  {http://arxiv.org/abs/1112.4423} {arXiv:1112.4423 [hep-ph]} \BibitemShut
  {NoStop}%
%%CITATION = 1112.4423;%%
\bibitem [{boe()}]{boer}%
  \BibitemOpen
  \href@noop {} {}\bibinfo {note} {We thank Daniel Boer for suggesting this
  form.}\BibitemShut {Stop}%
\end{thebibliography}%

\end{document}